\documentclass{aa}
\usepackage{graphicx}
\usepackage{tikz}
\usepackage[varg]{txfonts}
\usepackage{sansmath}
\usepackage{bm}
\usepackage{xcolor}
\usepackage[colorinlistoftodos]{todonotes}
\usepackage{bold-extra}
\usepackage{float}
\usepackage{soul}
\newcommand{\simlt}
{\,\hbox{\lower0.6ex\hbox{$\sim$}\llap{\raise0.6ex\hbox{$<$}}}\,}
\newcommand{\simgt}{\,\hbox{\lower0.6ex\hbox{$\sim$}\llap{\raise0.6ex\hbox{$>$}}}\,}

\authorrunning{Boyle et al.}
\titlerunning{Helium in Double-Detonation Models}

\begin{document}

\title{Helium in Double-Detonation Models of Type Ia Supernovae}
\author{Aoife Boyle\inst{1}{\thanks{E-mail:
aoife@mpa-garching.mpg.de (AB)}}
\and Stuart A. Sim\inst{2}
\and Stephan Hachinger\inst{3,4} 
\and Wolfgang Kerzendorf\inst{5}}
\institute{Max-Planck-Institut f{\"u}r Astrophysik, Karl-Schwarzschild-Strasse 1, 85748 Garching bei M{\"u}nchen, Germany \and
Astrophysics Research Centre, School of Mathematics and Physics, Queen's University Belfast, Belfast BT7 1NN, UK \and
Institut f\"ur Theoretische Physik und Astrophysik, Universit\"at W\"urzburg, Emil-Fischer-Str. 31, 97074 W\"urzburg, Germany
\and
Leibniz Supercomputing Centre, Boltzmannstr. 1, 85748 Garching bei M{\"u}nchen, Germany
\and
European Southern Observatory, Karl-Schwarzschild-Strasse 2, 85748 Garching bei M{\"u}nchen, Germany}

\abstract{
The ``double-detonation'' explosion model has been considered a candidate for explaining astrophysical transients with a wide range of luminosities. In this model, a carbon-oxygen white dwarf star explodes following detonation of a surface layer of helium. One potential signature of this explosion mechanism is the presence of unburned helium in the outer ejecta, left over from the surface helium layer. In this paper we present simple approximations to estimate the optical depths of important He~{\sc i} lines in the ejecta of double-detonation models. We use these approximations to compute synthetic spectra, including the He~{\sc i} lines, for double-detonation models obtained from hydrodynamical explosion simulations. Specifically, we focus on photospheric-phase predictions for the near-infrared 10830 \AA~and 2 $\mu$m lines of He~{\sc i}. 
We first consider a double detonation model with a
luminosity corresponding roughly to normal SNe~Ia. This model has a post-explosion unburned He mass of 0.03 $M_{\odot}$ and our calculations suggest that the 2 $\mu$m feature is expected to be very weak but that the 10830 \AA~feature may have modest opacity in the outer ejecta. Consequently, we suggest that a moderate-to-weak He~{\sc i} 10830 \AA~feature may be expected to form in double-detonation explosions at epochs around maximum light. However, the high velocities of unburned helium predicted by the model ($\sim 19,000$~km~s$^{-1}$) mean that the He~{\sc i} 10830 \AA~feature may be confused or blended with the C~{\sc i} 10690~\AA~line forming at lower velocities. 
We also present calculations for the He~{\sc i} 10830 \AA~and 2 $\mu$m lines for a lower mass (low luminosity) double detonation model, which has a post-explosion He mass of 0.077 $M_{\odot}$. In this case, both the He~{\sc i} features we consider are strong and can provide a clear observational signature of the double-detonation mechanism.}

\keywords{supernovae: general - white dwarfs - radiative transfer}

\maketitle

\section{Introduction}\label{Sec_Intro}

It is well-established that thermonuclear explosions of white dwarf (WD) stars may account for a variety of observed classes of astrophysical transient. One means to trigger such an explosion is via the detonation of a helium layer on the surface of a carbon-oxygen (CO) WD: if only the surface helium detonates, this may give rise to a relatively faint/fast transient \citep{bildsten2007}; alternatively, if the helium detonation triggers a secondary detonation of the underlying CO core, a much brighter explosion may occur, and this 
{\it double detonation} scenario
has been proposed as a possible model for both normal and subluminous Type Ia supernovae 
\citep[see e.g.][]{taam1980a, taam1980b, nomoto1980, nomoto1982, livne1990, woosleyweaver1994}. 
\cite{hoeflich1996} and \cite{nugent1997} both studied double-detonation models for systems with fairly massive (0.7 -- 0.9 $M_{\odot}$) CO cores and thick (0.15 -- 0.2 $M_{\odot}$) He outer layers, and compared results of radiative transfer calculations to observations of SNe~Ia. Overall, those studies disfavoured double-detonation models for normal SNe~Ia, owing to discrepancies in the light curve and spectral properties that could mostly be attributed to the presence of the helium-detonation ash in the outer ejecta of the models.
However, interest in double-detonation models was rekindled following suggestions by \citet[][ see also \citealt{shen2009,shen2010}]{bildsten2007} that surface helium detonation might be achievable in considerably lower mass helium shells. Subsequent explosion simulations have suggested that detonation in even low-mass helium layers may be sufficient to trigger secondary core detonation (\citealt{fink2010}, \citealt{sim2012}, \citealt{shen2014}; but see also \citealt{dunkley2013}), as required in the double-detonation model and -- moreover -- that the agreement between synthetic lightcurves/spectra and observations of normal SNe~Ia is somewhat improved if models with massive CO cores and low mass ($\simlt 0.05 M_{\odot}$) helium layers are invoked \citep{kromer2010,woosleykasen2011}.

\cite{kromer2010} and \cite{woosleykasen2011} extensively discuss the role of the heavy elements in the ash of the helium-shell detonation in shaping the light curves and spectra of double-detonation models with relatively low-mass outer helium layers. Both these studies draw attention to the role of the heavy elements in the outer ejecta affecting colours/spectral features and of surface radioactivity in shaping the light curve. Such effects could be powerful observational signatures of the double-detonation scenario. However, simulating the helium detonation, and the associated nucleosyntheis, is complex and the composition of the burning ash depends on a variety of currently unknown factors, including any potential pollution of the helium layer with e.g. carbon from the underlying core \citep{kromer2010, shen2010, shenmoore2014,townsley2012}. Here, we consider an alternative potential signature of double-detonation models: unburnt helium at high velocities. All the previous simulations have shown that significant masses of unburnt He should be present in the outer ejecta (e.g. across the \citealt{fink2010} series of models, approximately half of the inital shell mass is always ejected as unburned He, and \citealt{shenmoore2014} present calculations in which the ejected mass fraction of He for shell detonations around a 1.0~M$_{\odot}$ core range from $\sim 0.1 - 0.8$).
Thus a significant mass fraction of He in the outer ejecta is a robust prediction of double-detonation models, which should be considered when comparing their properties to observations.

In their studies of modern double-detonation models, neither \cite{kromer2010} nor \cite{woosleykasen2011} could fully address the question of whether helium spectral features should form in the models owing to approximations used in the atomic physics in both studies. However, in a recent study of a model for a helium surface detonation (but no secondary detonation of the underlying CO core), \cite{dessart2014} found that He~{\sc i} features are clearly expected when the full non-LTE (local thermodynamic equilibrium) physics is taken into consideration. The goal of this paper is to investigate whether contemporary double-detonation models could also be as clearly identified via He~{\sc i} features.

The challenge in modelling optical-NIR He~{\sc i} line features is that helium
excitation and ionisation rates in supernova ejecta are strongly affected by nonthermal electron collisions. Specifically, SNe~Ia are powered by the radioactive decay of $^{56}$Ni and it has been demonstrated in studies of type Ib and Ic supernovae (which are also powered by $^{56}$Ni decay) that the state of helium is strongly affected by fast electrons produced by $\gamma$-rays resulting from this radioactive decay \citep{chugai1987b, graham1988}. Helium is set apart from the other elements in its sensitivity to these effects because of the exceptionally large energy gap between the He~{\sc i} ground state and its first excited state (20 eV), and the metastability of its first two excited states. These properties lead to super-thermal electron collisions dominating many processes. A full non-LTE treatment of the state of helium takes into account the local generation of a population of fast electrons by Compton scattering / photoelectric absorption of $\gamma$-rays,
and accounts for these super-thermal electrons in calculating the ionisation and excitation state \citep{lucy1991, utrobin1996, mazzali1998, hachinger2012}. 
In this study we will make particular use of conclusions drawn by \cite{hachinger2012}, who made an extensive study that included nonthermal effects in calculating the state of He~{\sc i} in SN Ib/Ic simulations (see below). 

In Section~\ref{Sec_Spectral_Modelling}, we describe the approach we use to compute our synthetic spectra and the approximations we introduce to describe the helium level populations. In this study, we present results for two double-detonation models, which are introduced in Section~\ref{Sec_Supernova_Models}. We comment on numerical testing of our approach in Section~\ref{Sec_Testing} before presenting our calculated He~{\sc i} optical depths and synthetic spectra in Section~\ref{Sec_Results}. Our findings are discussed and summarised in Sections~\ref{Sec_Discussion} and \ref{Sec_Conclusions}. 

\section{Spectral Modelling}\label{Sec_Spectral_Modelling}

\subsection{Radiative Transfer Overview (\textsc{tardis} code)}\label{Sec_Codes}

We used \textsc{tardis} \citep{kerzendorf2014}, a Monte Carlo radiative transfer code, to model our spectra. The code operates in one dimension only and assumes that the supernova envelope is spherically symmetric. It works by simulating an artificial "photospheric" boundary deep inside the supernova envelope, which is assumed to emit a blackbody spectrum. The passage of these photons through the supernova ejecta and their interaction with matter is then simulated, and a convergent solution found for the radiation temperature, $T_{r}$, and electron density, $n_{e}$, which reproduces the desired luminosity. An output spectrum can then be generated.

\textsc{tardis} requires input specifying (i) a model for the supernova ejecta composition and density, (ii) the luminosity and time since explosion for which the spectrum is to be calculated, (iii) a set of atomic data (see Section \ref{Sec_Atomic}) and (iv) specification of a set of approximations to be used in computing the ionization/excitation state of the ejecta and the handling of matter-radiation interactions.

Each model consists of a number of concentric spherical shells, each being defined by inner and outer velocity boundaries (\textsc{tardis} assumes the whole ejecta are in homologous expansion and physical properties are assumed to be uniform within each shell). For our simulations, we used models for the density and composition based on previous hydrodynamical simulations (see Section \ref{Sec_Supernova_Models} for details). 
For each of the models we study, we have chosen to present calculations for three different epochs and we make use of light curve simulations to estimate appropriate luminosities to use as input for each calculation (see Section \ref{Sec_Supernova_Models}).

Table \ref{Table_Settings} gives a summary of important \textsc{tardis} numerical parameters and modes of operation used in the majority of our calculations.
These parameters are all explained in detail by \cite{kerzendorf2014}. 
Throughout this study, we have opted to use the simplest \textsc{tardis} treatment of bound-bound interactions (\texttt{scatter} mode, which treats all transitions via a resonant scattering approximation). As demonstrated by \citet[][see also \citealt{kerzendorf2014}]{lucy1999}, although simple, this approximation is generally adequate for modelling the optical spectra.

In the subsections below, we elaborate on the excitation/ionisation treatment used by \textsc{tardis} and the modifications that have been implemented for this study.

\begin{table}
\caption{Key modes and settings selected in the \textsc{tardis} configuration file for use in our simulations.}\label{Table_Settings}
\begin{tabular}{l l} \hline
Name & Setting\\
\hline
Ionisation mode & \texttt{nebular} \\
Excitation mode & \texttt{dilute-LTE} \\
Line interaction mode & \texttt{scatter} \\
Radiative rate mode & \texttt{dilute-blackbody} \\
Number of iterations & 30 \\
Number of packets\footnotemark[1] & $1.0\times 10^{6}$\\ 
\hline
\end{tabular}
\end{table}

\footnotetext[1]{There is one exception for this parameter. For the earliest spectrum generated for the low mass model (7.0 days after explosion), the number of packets was increased to $5.0\times 10^{6}$, due to relatively high optical depth.}

\subsection{\texttt{dilute-LTE} and \texttt{nebular} modes}

As described by \citet{kerzendorf2014}, \textsc{tardis} uses a variety of simple approximations to describe the ionization and excitation state in the ejecta, which are key ingredients of any spectrum synthesis calculation. For this study, we will make use of the \texttt{dilute-LTE} mode treatment of excitation and the \texttt{nebular} mode for ionization for most elements.

In \textsc{tardis} \texttt{dilute-LTE} mode \citep{kerzendorf2014}, the level populations of each ion are obtained from:

\begin{equation}\label{LTE_E}
n_{i,j,k} = W\frac{g_{i,j,k}}{Z_{i,j}}N_{i,j}\exp{\left( {-\frac{\epsilon_{i,j,k}}{kT_{r}}} \right)}
\end{equation}

\noindent where the subscripts $i$, $j$ and $k$ denote a specific element, ion and level, respectively. $n$ is a level population, $N$ is the total population of the corresponding ion, $Z$ is the partition function, $g$ represents the statistical weight of the level and $\epsilon$ is the energy of the level. The dilution factor \emph{W} is calculated during the simulation and approximately accounts for the dilution of the radiation field relative to a blackbody at the local radiation temperature \citep{mazzalilucy1993}.
\emph{W}
is excluded from Equation~\ref{LTE_E} when calculating ground state and metastable state populations. 

In \texttt{nebular} ionization mode, the relative ion populations are computed, following \cite{mazzalilucy1993}, via:

\begin{equation}\label{LTE_I}
\frac{N_{i,j+1}n_{e}}{N_{i,j}} = (W\left[\delta(\zeta_{i,j} + W(1-\zeta_{i,j}))\right]\left(\frac{T_{e}}{T_{r}}\right)^{\frac{1}{2}}\Phi_{i,j,T_{r}}
\end{equation}

\noindent where:

\begin{equation}
\Phi_{i,j,T_{r}} = \frac{2Z_{i,j+1}(T_{r})}{Z_{i,j}(T_{r})}\left(\frac{2\pi m_{e}kT_{r}}{h^{2}}\right)^{\frac{3}{2}}\exp{\left({-\frac{\chi_{i,j}}{kT_{r}}}\right)}
\end{equation}

\noindent $n_{e}$ is the electron density and $\chi_{i,j}$ is the ionisation potential. The dilution factor \emph{W} is defined as before and the factor $\zeta_{i,j}$ gives the fraction of recombinations that go directly to the ground state \citep{mazzalilucy1993}. The parameter $\delta$ approximately accounts for line blanketing in the blue part of the spectrum, and adjusts the ionization balance accordingly \citep[for a more complete explanation of the $\delta$ factor, see, as an example,][]{mazzalilucy1993}.

Throughout this study, the \texttt{dilute-LTE} and \texttt{nebular} modes will be used for all elements {\it apart} from He (see below). 

\begin{center}

\begin{figure*}

\includegraphics[width=\textwidth]{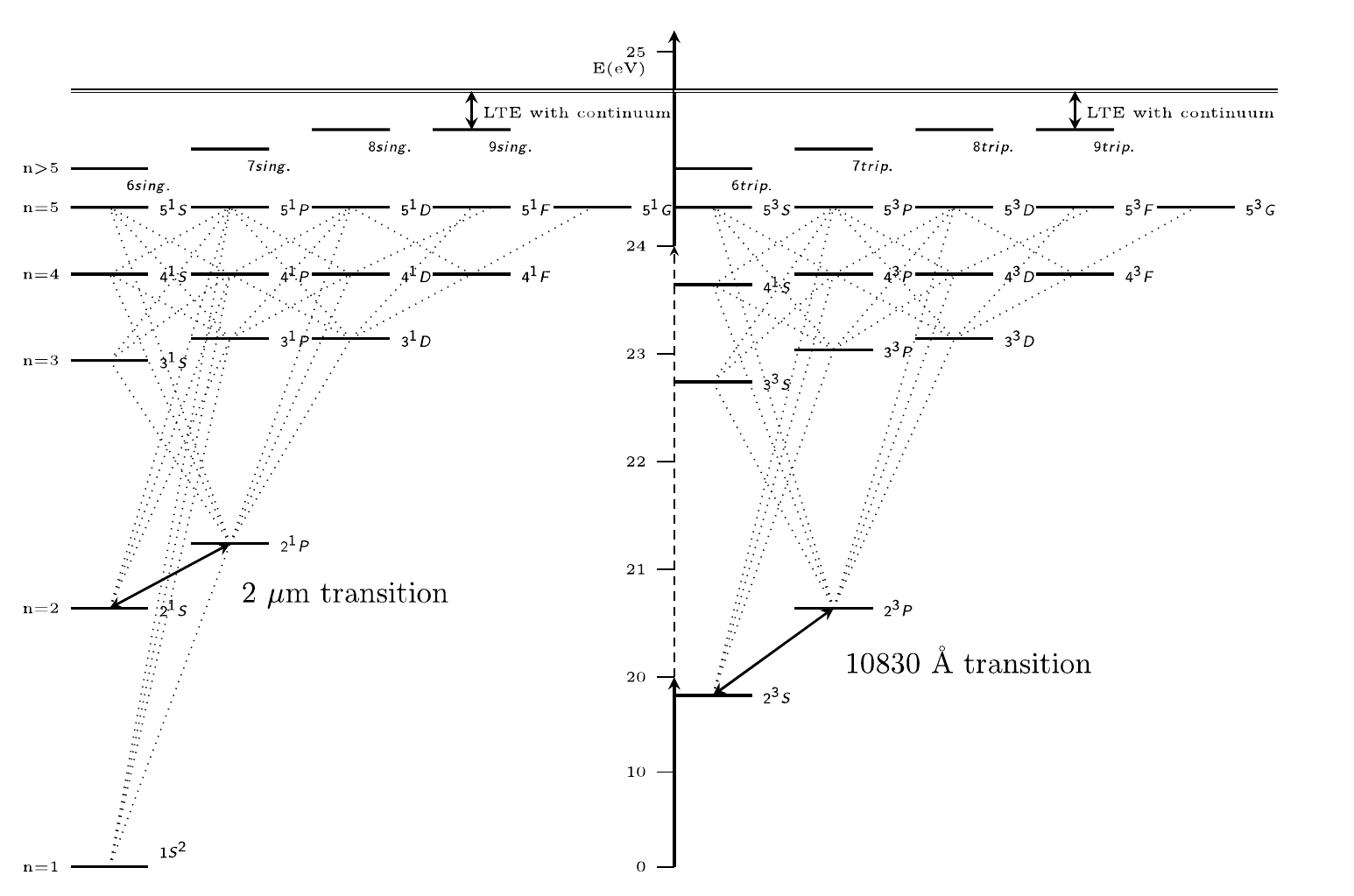}

\caption{Left: Helium I energy levels in the parahelium (S=0) state. The levels $6\leq n\leq 9$ are grouped into one singlet and one triplet level each and these are denoted as $n_{sing.}$ For clarity, the energy scale changes as the energy levels become closer at E=20 eV and E=24 eV, emphasised here by the change in line style of the energy axis at these points. The dotted lines show electric dipole permitted transitions. Although not included (to prevent overcrowding of the diagram), the $6\leq n\leq 9$ singlet levels possess allowed transitions with all of the other levels. Right: Helium I energy levels in the orthohelium (S=1) state. Note that the lowest excited energy states of both regimes are metastable.}
\end{figure*}

\end{center}

\subsection{New Approximation for the He State (\texttt{recomb-NLTE})}\label{Sec_New_Approx}

\cite{hachinger2012} noted that, in their simulations for type Ib/Ic supernova models, the He~{\sc i} excited level populations
were found to be more closely coupled to the He~{\sc ii} ground state than the He~{\sc i} ground state. They explained this as being the result of the dominance of the He~{\sc ii} ion population (due to non-thermal ionisation effects), and specific atomic properties of the He atom, outlined in Section \ref{Sec_Intro}.

Based on these findings, we have developed an analytic approximation (implemented as a new \texttt{recomb-NLTE} mode in {\sc tardis}) for the helium level populations that does not involve full numerical NLTE calculations. 
The value of such an approximation is to give insight into the key components of the underlying atomic physics by testing a simplified model: the results of comparisons with full NLTE treatment are provided in Section \ref{Sec_Results}. The approximation also allows calculations to be performed more efficiently than by coupling radiative transfer codes to a full NLTE solver.

In the \texttt{recomb-NLTE} approximation, the He~{\sc ii} excited state populations and He~{\sc iii} ground state population are calculated relative to the He~{\sc ii} ground state population using Equations \ref{LTE_E} and \ref{LTE_I}. The He~{\sc i} ground state population is assumed to be negligible, 
and for the majority of He~{\sc i} excited states the populations are treated as though in dilute LTE with the He~{\sc ii} ground state using the equation:

\begin{equation}\label{LTE_D}
\frac{n_{2,0,k}}{n_{2,1,0}n_{e}} = \frac{1}{W}\frac{g_{2,0,k}}{2g_{2,1,0}}\left(\frac{h^{2}}{2\pi m_{e}kT_{r}}\right)^{\frac{3}{2}}\exp{\left( {\frac{\chi_{2,1}-\epsilon_{2,0,k}}{kT_{r}}} \right)}
\end{equation}
We include the factor of $\frac{1}{W}$ to approximately account for the dilution of the ionizing radiation field, which leads to an 
increase in the helium excited state populations relative to LTE. 

The two lowest excited states of He~{\sc i}, the $2^{3}$S and $2^{1}$S states, are treated slightly differently. These two states are very significant because of their roles in the creation of the helium 10830 \AA~line and 2 $\mu$m line, respectively (see Section \ref{Sec_Atomic}). For these levels, an extra factor $\frac{1}{W}$ is included, as these states are metastable and more strongly populated by the cascade from higher levels than by direct interactions with the He~{\sc ii} ground state. Therefore, these two states are populated according to:

\begin{equation}\label{LTE_M}
\frac{n_{2,0,k}}{n_{2,1,0}n_{e}} = \frac{1}{W^2}\frac{g_{2,0,k}}{2g_{2,1,0}}\left(\frac{h^{2}}{2\pi m_{e}kT_{r}}\right)^{\frac{3}{2}}\exp{\left( {\frac{\chi_{2,1}-\epsilon_{2,0,k}}{kT_{r}}} \right)}
\end{equation}

The combination of these equations (Eqs. \ref{LTE_E}, \ref{LTE_I}, \ref{LTE_D}, \ref{LTE_M}) allows calculation of the populations of all helium ions/levels relative to the He~{\sc ii} ground state population, assuming the He~{\sc i} ground state population is negligible (compared to the total He population). In our \texttt{recomb-NLTE} mode, these calculations are made, and the  populations are then normalised to the total number of helium atoms in the relevant model zone. 

Our approximation can only be valid while the He~{\sc ii} ion population is dominant. For the combination of models and epochs we will consider, however, this is likely to be a reasonable assumption that is consistent with previous work. 
Specifically, 
\cite{dessart2014} found that the helium in their simulations of ``.Ia'' supernovae (accounting for NLTE effects on the state of helium) remained strongly ionised for a significant period of time after maximum light. Similarly, \cite{hachinger2012} found in their SN Ib/Ic simulations that helium remained ionised to a significant degree until at least a few days after maximum light, but not completely ionised.
We specifically test the validity of this assumption in our simulations by comparison to full NLTE level population calculations in Section~\ref{Sec_Results}.

We stress that the approximation described here is specific to helium and is only used for that element in our calculations. All other elements continue to be treated using our standard \texttt{dilute-LTE}/\texttt{nebular} approach (see Section~\ref{Sec_Codes}).

\subsection{Atomic Models}\label{Sec_Atomic}

For all elements (\textsc{tardis} considers those with atomic numbers $1\leq Z\leq 30$) apart from helium, we used 
the same atomic data described by \cite{kerzendorf2014}.
For helium, we adopted the atomic energy level and transition line data
\cite{hachinger2012} \citep[originally derived from data from the NIST Atomic Spectra Database, see][]{ralchenko2005}. This model consists of 29 He~{\sc i} levels with principal quantum numbers up to $n=5$, and beyond this value groups levels so that there is one singlet and one triplet state for each value of $n$ up to a final quantum number $n=9$ (see Figure~2).

\section{Supernova Models}\label{Sec_Supernova_Models}

We have investigated two double-detonation models, which we designate the HM (high-mass) and LM (low-mass) models. Our HM model is intended to be representative of recent double-detonation models for normal-luminosity SNe Ia \citep{fink2010,kromer2010}, while our LM model might be considered appropriate for a sub-luminous thermonuclear explosion \citep{sim2012}. Both models were obtained from two-dimensional hydrodynamical simulations in which detonation of a surface He layer was triggered at a point \citep[for details, see][]{fink2010,sim2012}. The He detonation then propagates laterally through the He layer, driving a shock into the underlying CO core. Detonation of the CO core is then triggered at the shock convergence point.

\begin{figure}

\includegraphics[width=0.5\textwidth]{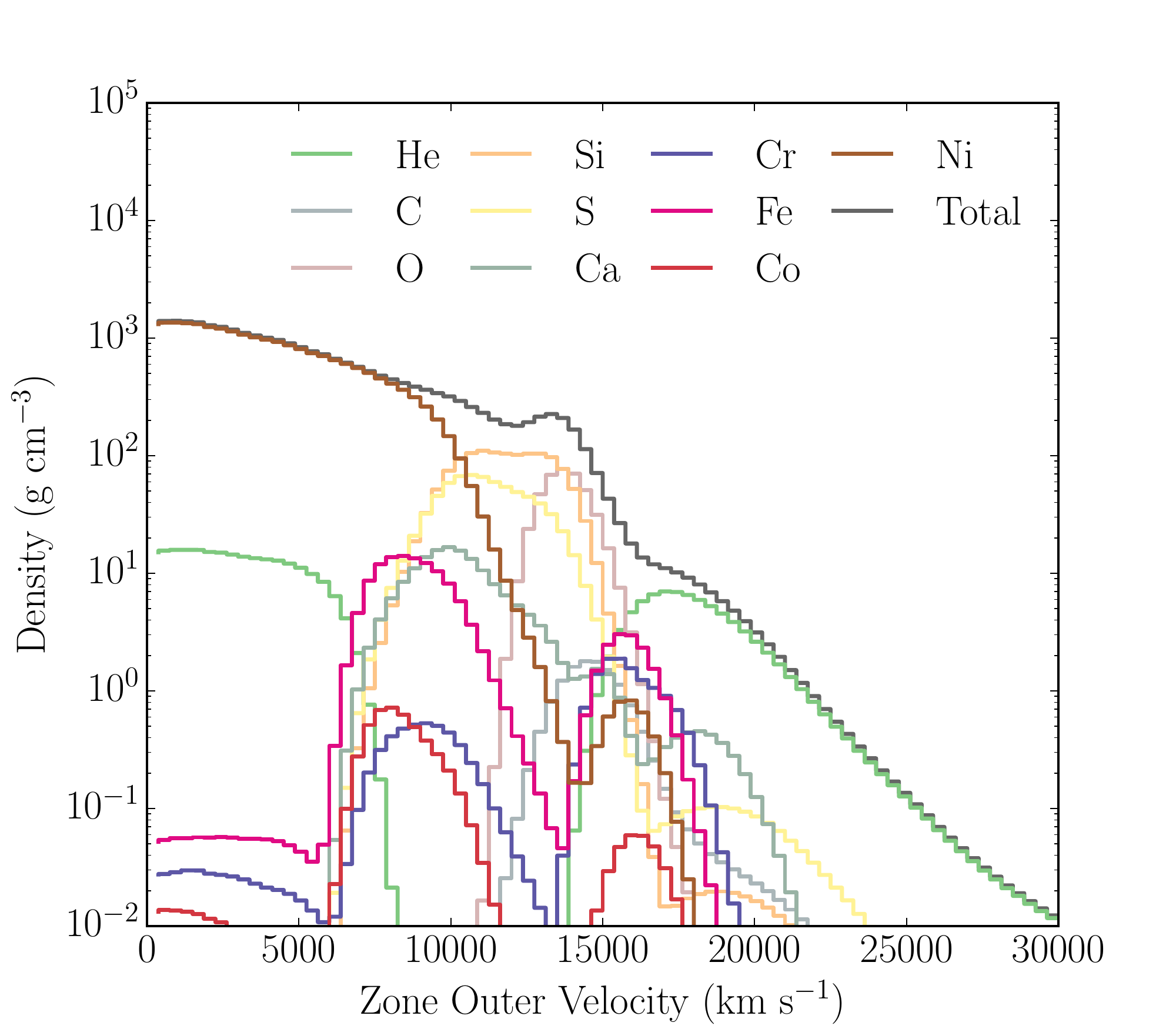}
\caption{Densities of the high-mass (HM) double-detonation model (most abundant elements only) around 7 seconds after explosion.}\label{fig_abundances_hm}

\end{figure}

The HM model is based on an equatorial slice through the 2D ``Model 3'' of \cite{fink2010}, and was also studied by \cite{kromer2010}. The simulation was based on a progenitor that had a CO core mass of 1.025 $M\odot$, and a helium shell mass of 0.055 $M_{\odot}$. After explosion, the model has 
0.03 $M_{\odot}$ of unburned helium remaining. 
The post-explosion composition and structure of the HM model are shown in Figure \ref{fig_abundances_hm}. 

\begin{figure}

\includegraphics[width=0.5\textwidth]{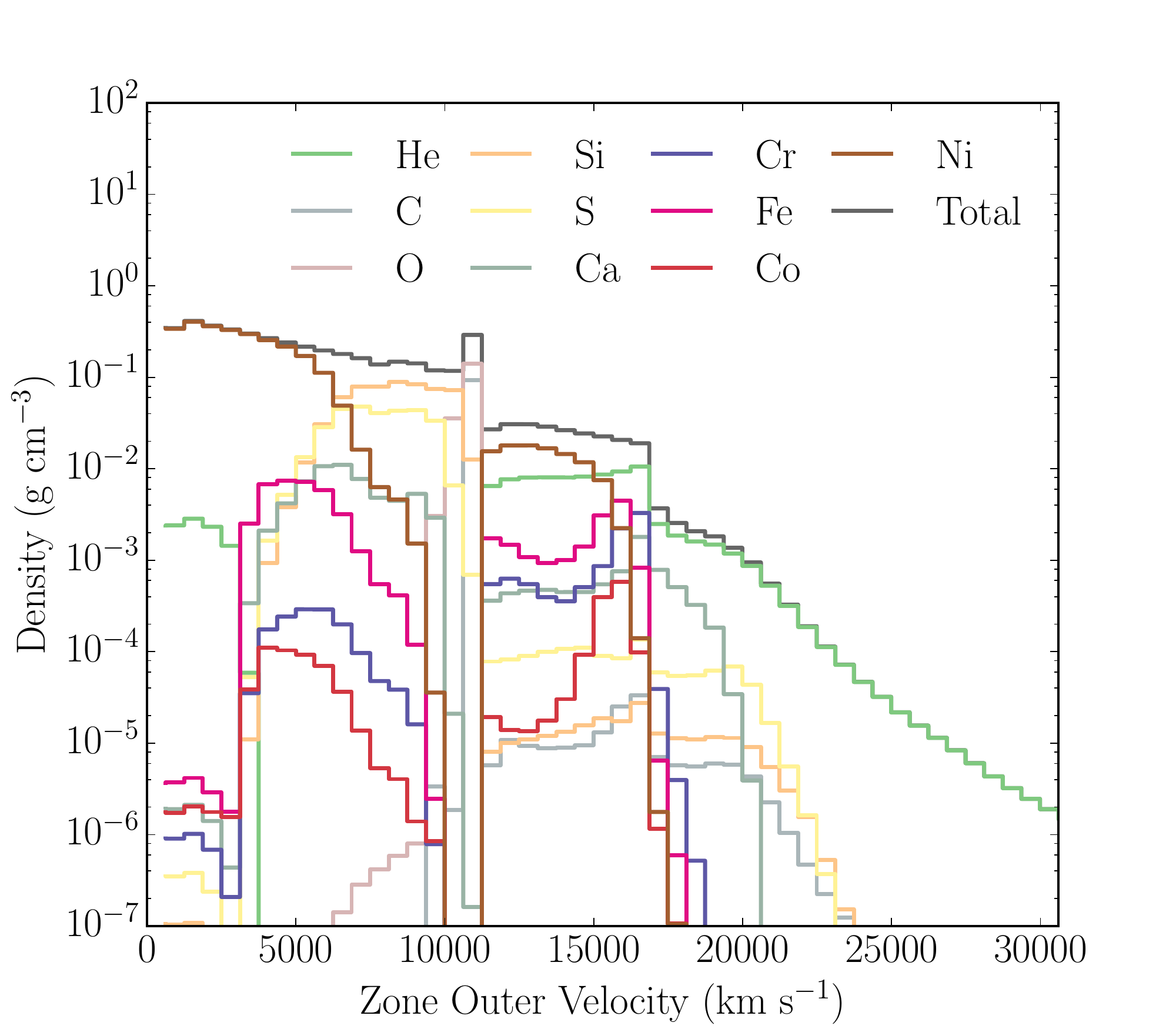}
\caption{Densities of the low-mass (LM) double-detonation model (most abundant elements only) about 100 seconds after explosion.}\label{fig_abundances_lm}

\end{figure}

The LM model was obtained from an equatorial slice of the ``CSDD-S'' model from \cite{sim2012}. Before explosion, this model has a CO core mass of 0.58 $M_{\odot}$ and a He shell mass of 0.21 $M_{\odot}$. After explosion, the remaining mass of helium was 
0.077 $M_{\odot}$, and the structure is shown in Figure \ref{fig_abundances_lm}. 
We note that, like the HM model, this model is based on assuming that a secondary CO detonation is triggered via shock convergence in the core \citep[see][]{sim2012}, although this mechanism is more difficult to realise in the low mass regime \citep[see][]{shen2014}.
It is also possible that double-detonations can occur via an {\it edge-lit} mechanism, in which the initial He detonation triggers a CO detonation from the surface of the CO core \citep{nomoto1982, livneglasner1990, forcada2006, forcada2007}. If realised, this different secondary ignition mechanism would be expected to lead to quantitative changes in the ejecta structure (see fig. 2 of Sim et al. 2012) but has only a modest effect on the mass of unburned helium in the outer ejecta (see table 3 of Sim et al. 2012).

Both of our models consist of 80 radial shells and both are assumed to be in homologous expansion for all times later than the final times of the hydrodynamical calculations on which they are based.

For our {\sc tardis} simulations, 
we placed the outer-velocity boundaries of our computational domain at $\sim 30,000$ km~s$^{-1}$, sufficiently large to include the He-rich high-velocity ejecta.

$^{56}$Ni dominates the inner ejecta in both models. These iron-rich inner layers can be expected to remain relatively optically thick and so we opted to place the inner velocity boundaries for our simulations approximately at the outer boundary of the iron-rich core. Sensitivity to the choice of inner boundary velocity was tested, and it was found that moving a few shells inwards or outwards made no qualitative difference to the output spectra or the conclusions we draw from them.

For both models we have computed \textsc{tardis} spectra for three epochs.
To estimate appropriate luminosities to adopt for  each \textsc{tardis} simulation we used 
\textsc{artis} \citep{kromer2009,sim2007} to calculate theoretical light curves for our models.
These light curve calculations were also used to define our time epochs (maximum light, a week before maximum light, and a week after). In addition, \textsc{artis} provided the $\gamma$-ray energy deposition rates ($H_\gamma$)
used to account for the effect of nonthermal collisions on the state of helium. These rates are not directly needed in our \textsc{tardis} calculations but 
are used as input to the statistical equilibrium solver of \cite{hachinger2012} to which we will compare our results (see below). We note that, although \textsc{artis} does provide time series spectra, we have opted to use \textsc{tardis} (instead of \textsc{artis}) for our detailed studies of helium line formation for its efficiency, and because its modular structure made it straightforward to modify and test our \texttt{recomb-NLTE} approximation.

Table \ref{Table_Model_Summary} summarises the input model parameters used for the spectrum synthesis calcualtions that will be presented in Section \ref{Sec_Results}.
All of the spectra that we show have been processed using a Savitzky-Golay filter \citep{savitzkygolay}.

\begin{table}
\caption{Simulation parameters at the various epochs for both models: time (\emph{t}) relative to maximum light (ML), time since explosion (\emph{$t_{exp}$}), luminosity ($L$) and velocity range (\emph{v} range).
}\label{Table_Model_Summary}
\begin{tabular}{l  c  c  c } \hline
{\emph{t}} & \emph{$t_{exp}$} & {\emph{L}} & {\emph{v} range}\\
(days) & (days) & ($L\odot$) & (km s$^{-1}$)\\ 
\hline
HM Model\\
ML - 7 days & 12 & 9.34 &9,400 - 30,000\\
ML & 19 & 9.51 &9,400 - 30,000\\
ML + 7 days & 26 & 9.32 &9,400 - 30,000\\
\hline
LM Model\\
ML - 7 days & 7 & 8.87 &6,250 - 30,600\\
ML & 14 & 8.97 &6,250 - 30,600\\
ML + 7 days & 21 & 8.91 &6,250 - 30,600\\
\hline

\end{tabular}

\end{table}

\section{Testing}\label{Sec_Testing}

\begin{figure}
\includegraphics[width=0.5\textwidth]{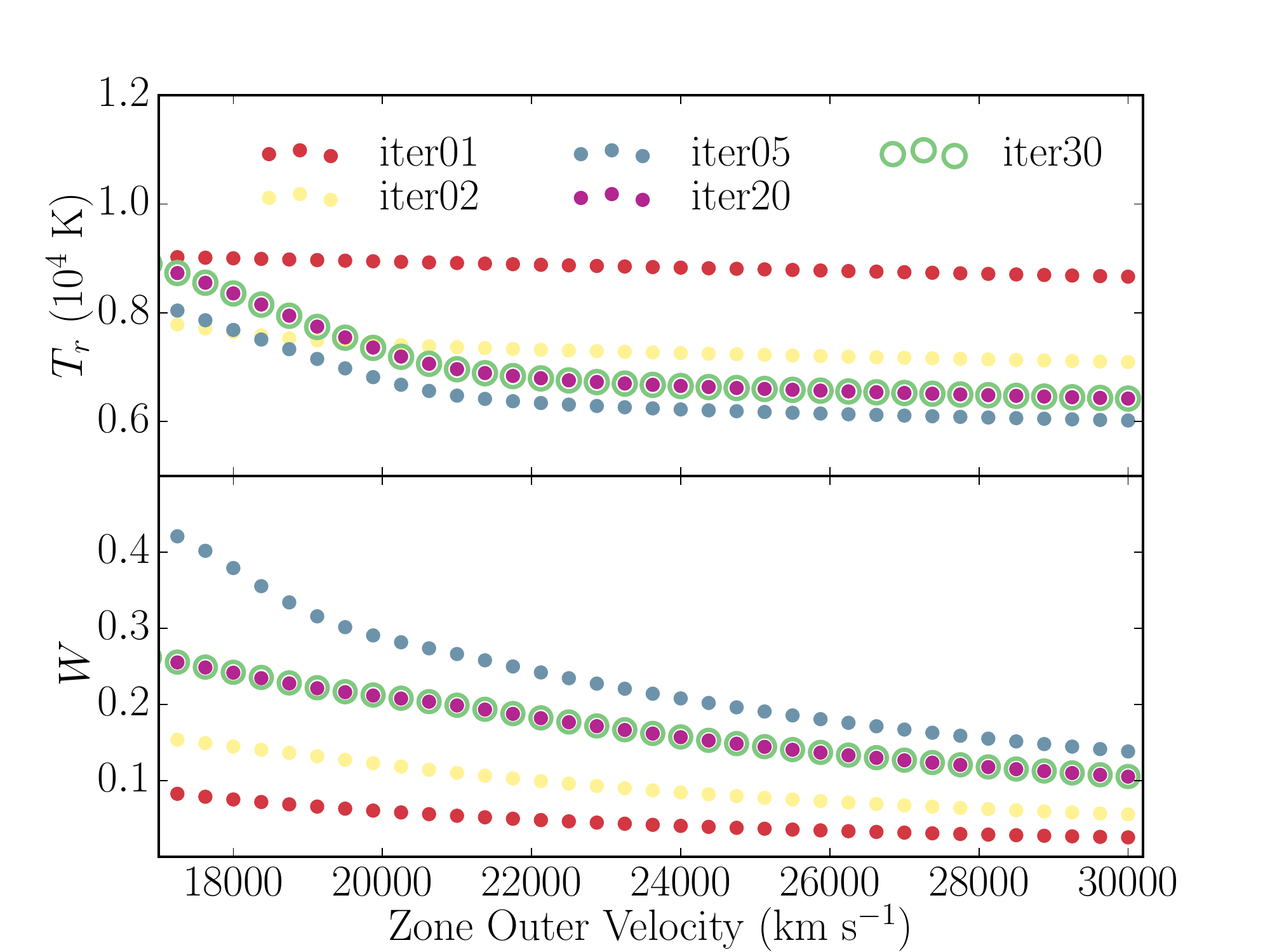}
\caption{The convergence processes for $T_{r}$ and $W$ for the HM model at maximum light (19.0 days after explosion), over 30 iterations when using the \texttt{recomb-NLTE} treatment. A convergent solution is found for all three parameters by the 20th iteration.
}\label{fig_convergence}
\end{figure}

A number of tests were carried out to check the robustness of our calculations. Convergence was tested by monitoring the radiation temperature $T_{r}$ and the dilution factor $W$ in each shell over 30 iterations of the code. These results are presented graphically in Figure \ref{fig_convergence}. In each case, it is clear that after 20 iterations, each of these values becomes effectively constant.

Throughout this study we have continued to follow the assumptions of \cite{mazzalilucy1993} and \cite{hachinger2012} in specifying the electron temperature $T_{e}$ by:

\begin{equation}
T_{e} = 0.9\times T_{r}
\end{equation}
We tested the sensitivity of our results to this assumption, and found that varying the value of $T_{e}/T_{r}$ made effectively no qualitative difference to the helium features in the final spectra. A variety of inner velocity boundary values were also sampled to ensure that the resulting spectra were not qualitatively sensitive to this choice.

\section{Results}\label{Sec_Results}

Our analysis focuses on the He~{\sc i} 10830 \AA~($2^{3}$S -- $2^{3}$P) and 2~$\mu$m He~{\sc i} ($2^{1}$S -- $2^{1}$P) transitions.
These are expected to be the most easily observed He~{\sc i} features due to their relatively high oscillator strengths and high populations in the metastable 2S states. In addition, identification of these features in observed spectra can be relatively robust since they lie in spectral regions that are less affected by strong line blending compared to higher excitation He~{\sc i} transitions in the bluer parts of the optical.

\subsection{High Mass Model}

We have carried out six \textsc{tardis} simulations for the HM model: for each of the three epochs listed in Table~\ref{Table_Model_Summary} we made two calculations, one with our new \texttt{recomb-NLTE} mode activated for helium and one without.

\begin{figure}
\includegraphics[width=0.5\textwidth]{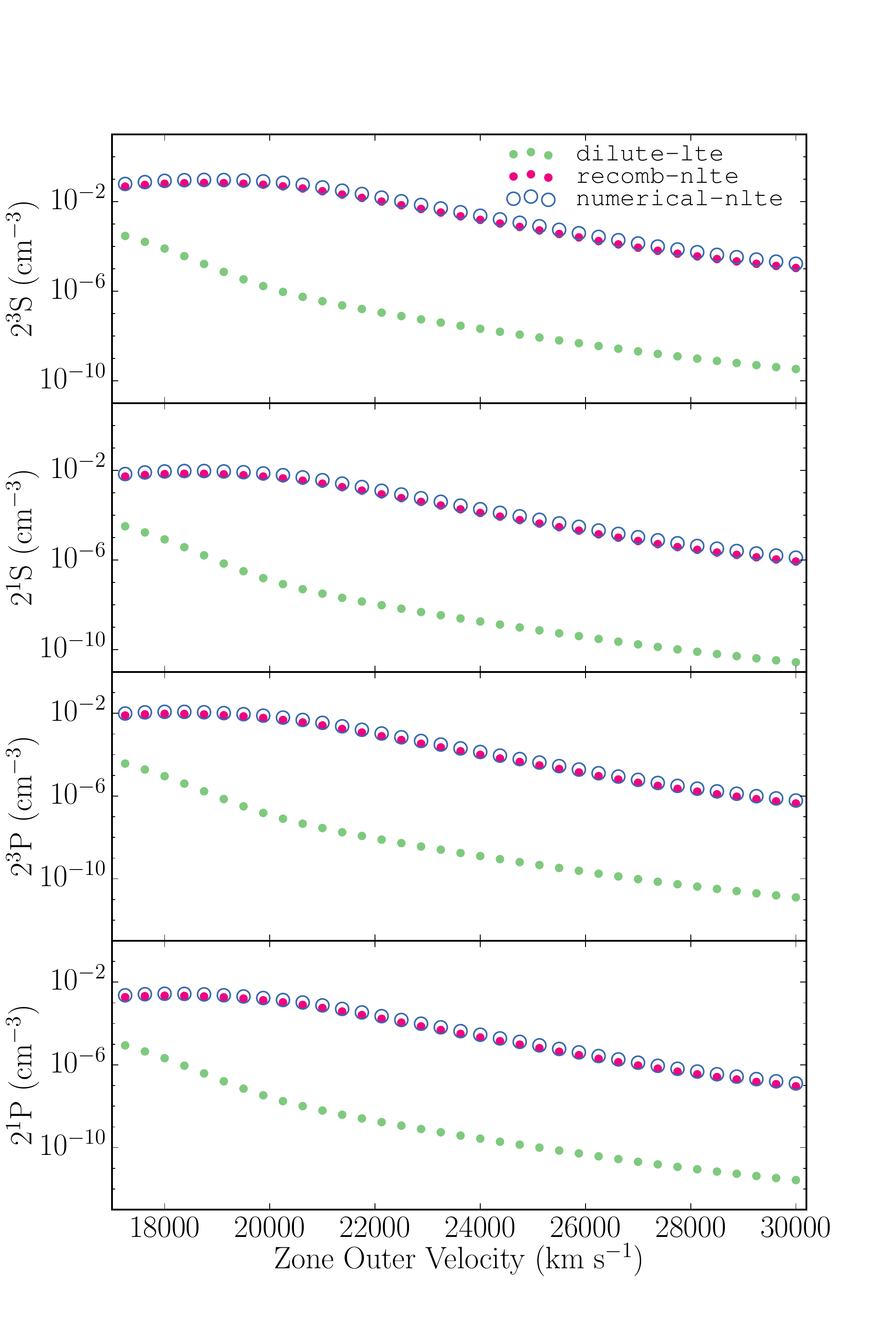}
\caption{Level populations for the relevant lines for the HM model at maximum light as a function of velocity. \texttt{numerical-nlte} refers to calculations made with the full statistical equilibrium solver module of Hachinger et al. (2012, see text).}\label{fig_level_populations_hm}
\end{figure}

\begin{figure}
\includegraphics[width=0.5\textwidth]{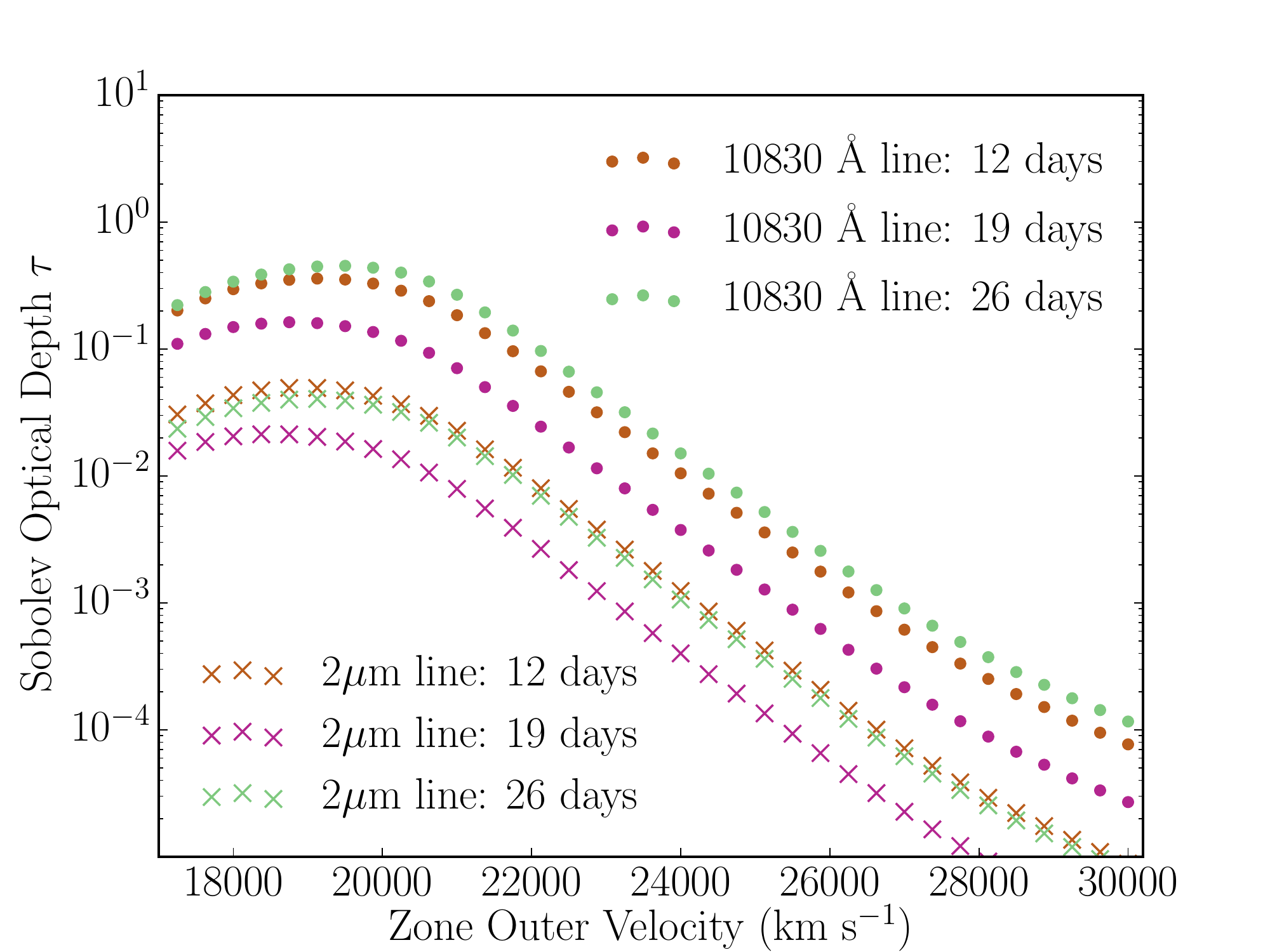}
\caption{Values of $\tau_{sob}$ for the 2 $\mu$m line and the 10830 \AA~line for the He-rich zones of the HM model, at all 3 epochs, produced using the \texttt{numerical-NLTE} treatment.}\label{fig_tau_sobolevs_hm}
\end{figure}

\begin{figure*}

\includegraphics[width=\textwidth, height=15cm]{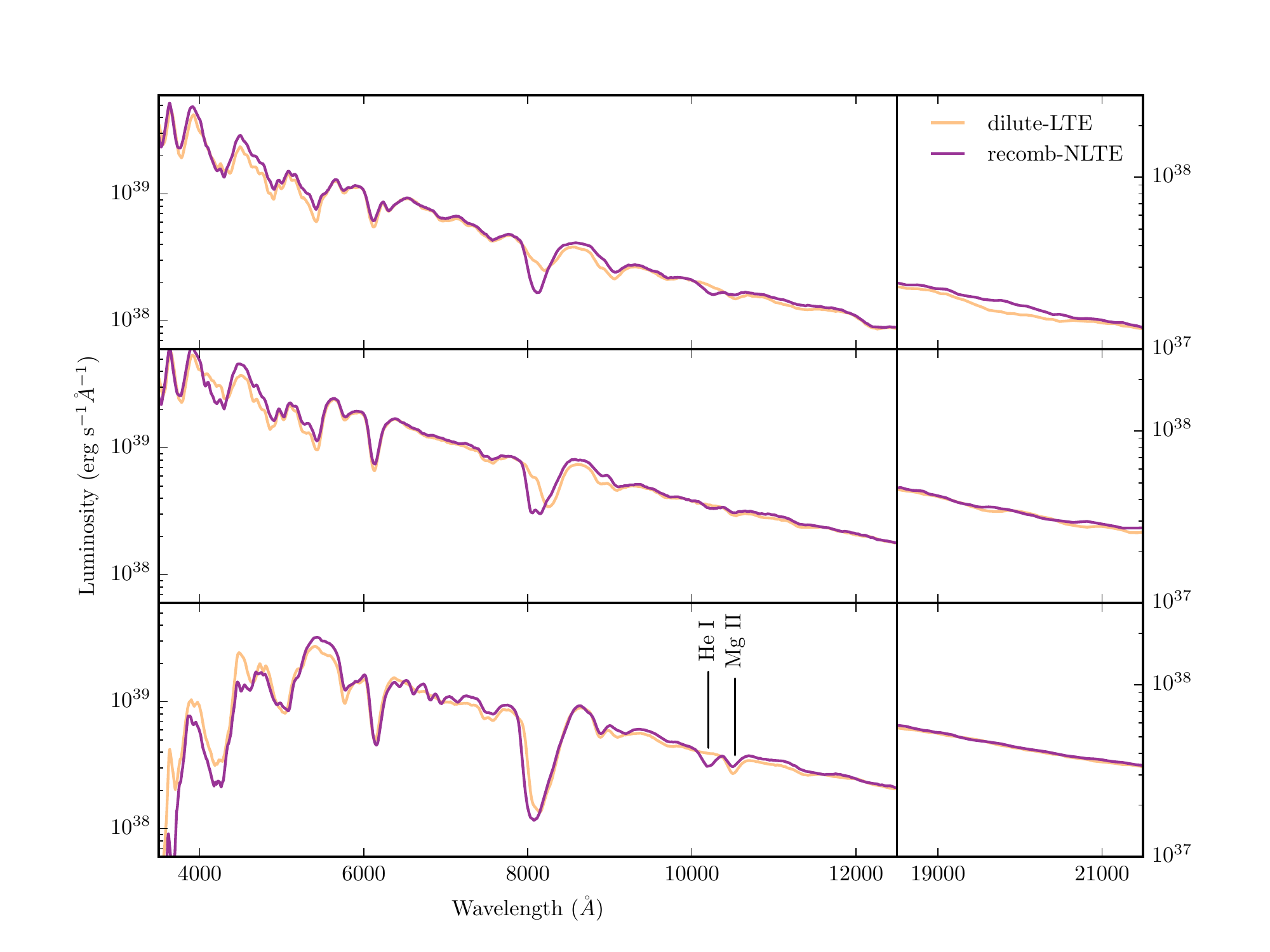}
\caption{Spectra obtained for the HM model at epochs 12, 19 and 26 days after explosion. All of the spectra in this paper have had noise removed using a Savitzky-Golay filter \citep{savitzkygolay}. The pink spectra are those obtained with the original \textsc{tardis} code without taking NLTE effects into account. The green spectra are those obtained using our new approximation for imitating NLTE effects. In this case, the 10830 line appears weakly at late times, as highlighted in pink. The Mg {\sc ii} 10927~\AA-feature that could be mistaken for the He 10830 \AA~line is also highlighted in yellow.}\label{fig_spectra_hm}

\end{figure*}

\subsubsection{Helium level populations and optical depths}
\label{Sect_levelpop_hm}

In Figure \ref{fig_level_populations_hm} we show the level populations obtained by \textsc{tardis} for the HM model at maximum light, as a function of zone velocity, for the four levels involved in the two transitions on which we focus. In addition to showing results from our two \textsc{tardis} simulations, we also show level populations computed by solving a full NLTE system of statistical equilibrium equations for helium.
These calculations were made using the statistical equilibrium solver described by \cite{hachinger2012}: one calculation was made for each zone in the model using the appropriate composition / density and adopting the local radiation field model obtained by \textsc{tardis} and the non-thermal heating rate obtained from \textsc{artis}.
Figure \ref{fig_level_populations_hm} shows that the \texttt{recomb-NLTE} analytic treatment replicates the full NLTE statistical equilibrium results well under these conditions: in the zones with significant densities of He, the approximation predicts level populations within $\approx$ 30 \% of the full numerical NLTE values. This is a very dramatic improvement over the \texttt{dilute-LTE} treatment, which, as expected, deviates by orders of magnitude.

Our calculations of the Sobolev optical depths of all He~{\sc i} lines show that the strongest He~{\sc i} feature expected in the optical/IR spectra is the 10830 \AA~line. 
The calculated Sobolev optical depths of the 2 $\mu$m line and the 10830 \AA~line are shown for the He-rich shells of the HM model in Figure \ref{fig_tau_sobolevs_hm} for the three epochs considered. Based on these optical-depth calculations, it is clear that absorption by 
the He~{\sc i}
2~$\mu$m line will be very weak at all the epochs considered; however, although $\tau < 1$, some opacity is expected in the 10830~\AA~line ($\tau \sim 0.5$ at the post-maximum epoch). 
Although the three epochs we have considered span a factor of three in time since explosion, the maximum value of the optical depth (occurring around 19,000 km~s$^{-1}$) is fairly similar (with a factor of $\sim 3$) at all three epochs for both lines ($\tau \sim 0.15 - 0.5$ for the 10830 \AA~line and $\tau \sim 0.02 - 0.05$ for the 2 $\mu$m line). In detail, however, the predicted evolution is complicated (and likely to be strongly model dependent): the calculated optical depth decreases slightly between 12 and 19 days but then rises again by 26 days. This is a consequence of the interplay between the continuous reduction of ejecta densities and increase of velocity gradients with time (direct consequences of homologous expansion), and variations in the radiation temperature.

\subsubsection{Helium spectral features}

Our calculated spectra for the HM model at the three epochs we consider are shown in Figure \ref{fig_spectra_hm} (optical to J-band region in the left panel and window around the 2~$\mu$m region in the right panel). For comparison, we show calculations with and without activating our new \texttt{recomb-NLTE} mode for helium.

As anticipated from Figure~\ref{fig_tau_sobolevs_hm}, The He~{\sc i} 2 $\mu$m feature does not show up in the spectra for this model, regardless of the helium treatment used. However, the helium 10830 \AA~line does appear, albeit relatively weakly. The feature is strongest at the latest epoch we consider and is clearly separated from the adjacent Mg~{\sc ii} features around 10920 \AA.

It is interesting to note that adopting the \texttt{recomb-NLTE} mode for helium affects not only the He~{\sc i} line but also affects the shapes and strengths of other features, for example the high-velocity blue wing of the Ca~{\sc ii} NIR triplet around 8100 \AA. This will be discussed further in Section~\ref{Sec_Discuss_Other}, below.

\subsection{Low Mass Model}

\begin{figure}
\includegraphics[width=0.5\textwidth]{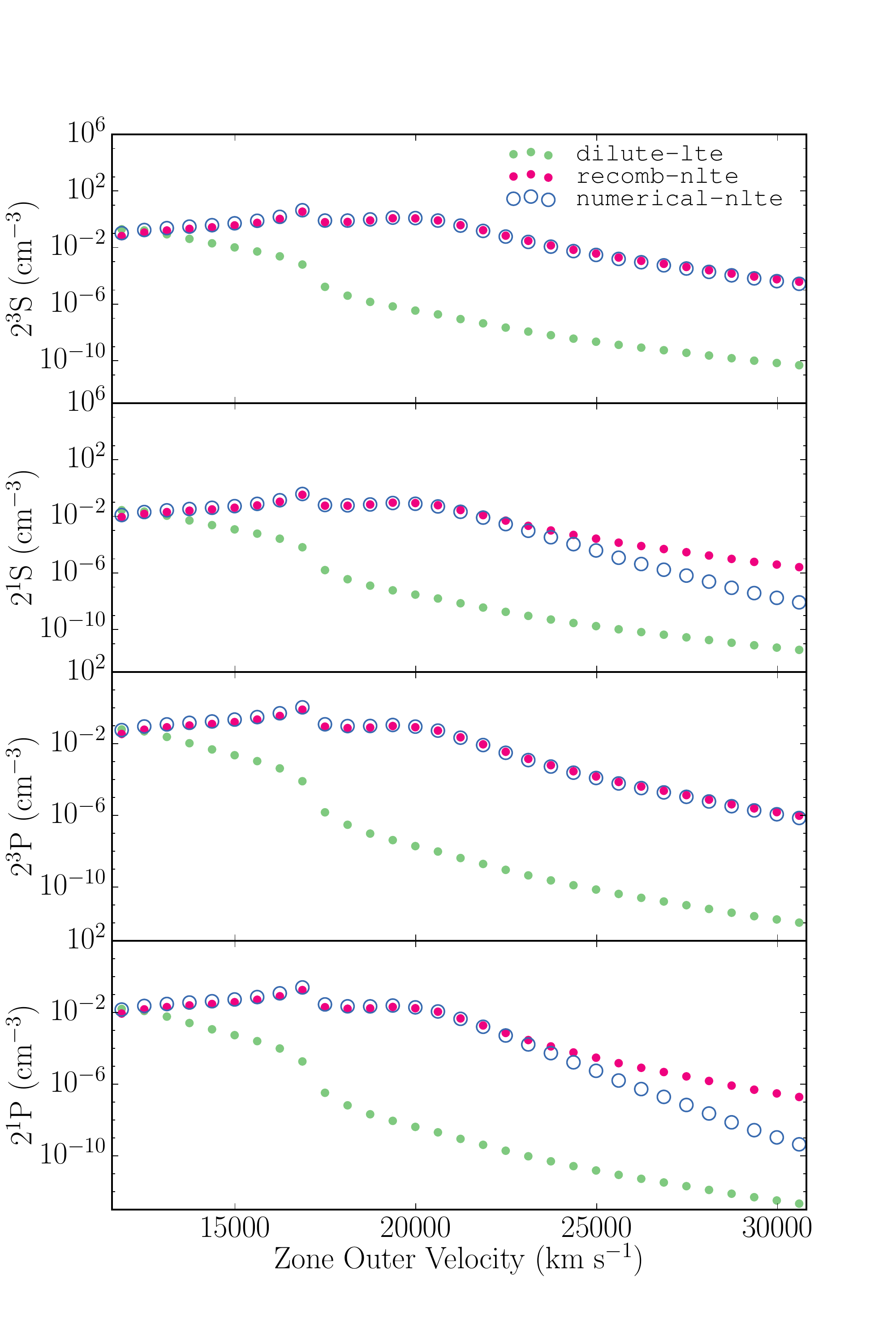}
\caption{As Fig.~\ref{fig_level_populations_hm} but for the LM model at maximum light.}\label{fig_level_populations_lm}
\end{figure}

\begin{figure}
\includegraphics[width=0.5\textwidth]{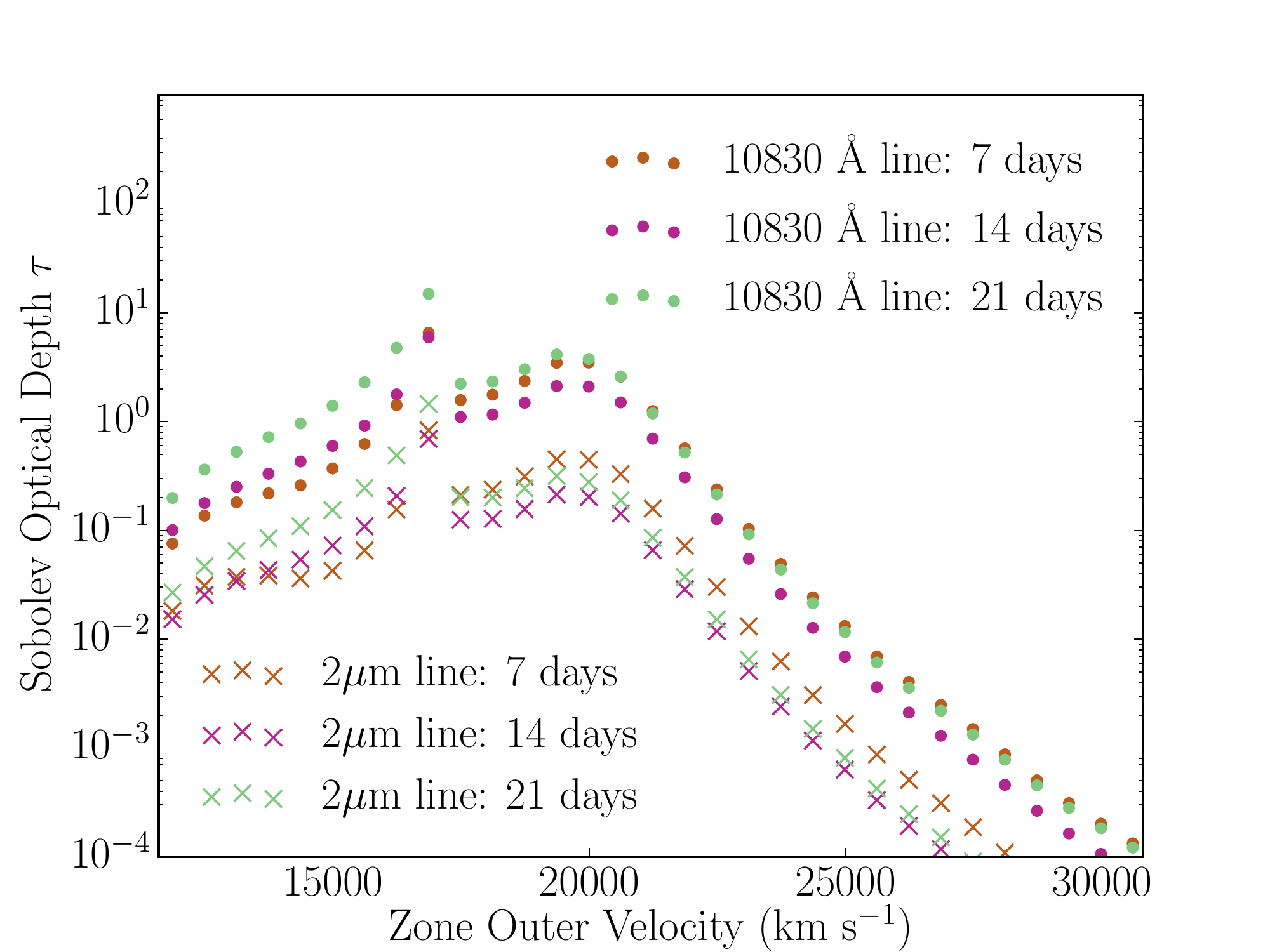}
\caption{As Fig~\ref{fig_tau_sobolevs_hm} but for the LM model.}\label{fig_tau_sobolevs_lm}
\end{figure}

Level populations and Sobolev optical depths for the LM model are shown in 
Figures~\ref{fig_level_populations_lm} and Figure \ref{fig_tau_sobolevs_lm}. As for the HM model, comparison of level populations suggests that the \texttt{recomb-NLTE} mode implemented in \textsc{tardis} provides a good approximation to the most relevant He~{\sc i} level populations throughout most of the outer ejecta. We do note in this case that the singlet states (2$^1$S and 2$^1$P) are clearly overestimated by the \texttt{recomb-NLTE} mode at the highest velocities ($> 25,000$ km~s$^{-1}$). However, it can be seen from Figure \ref{fig_tau_sobolevs_lm} that the optical depths in these outermost zones is low and thus the synthetic spectra will not be affected.

In contrast to the HM model, the optical depths for the LM model are sufficiently high that both the 2~$\mu$m and 10830~\AA~lines can be expected to be relatively strong. The optical depth is particularly high in the zones around 17,000~km~s$^{-1}$, just above the maximum velocity of the iron-group material that was synthesised in the the helium shell detonation (see Figure~\ref{fig_abundances_lm}). The relatively high optical depths in the LM model are primarily a consequence of the He density, the peak value of which is roughly 10 times higher than in the HM model (although the lower luminosity and temperature of the model do also enhance the He~{\sc i} populations compared to the HM case). As for the HM model, the optical depths are relatively similar across the wide range of epochs considered; again this is consequence of the interplay between homologous expansion (i.e. diluting density and increasing velocity gradient) and temperature evolution.

\begin{figure*}

\includegraphics[width=\textwidth, height=15cm]{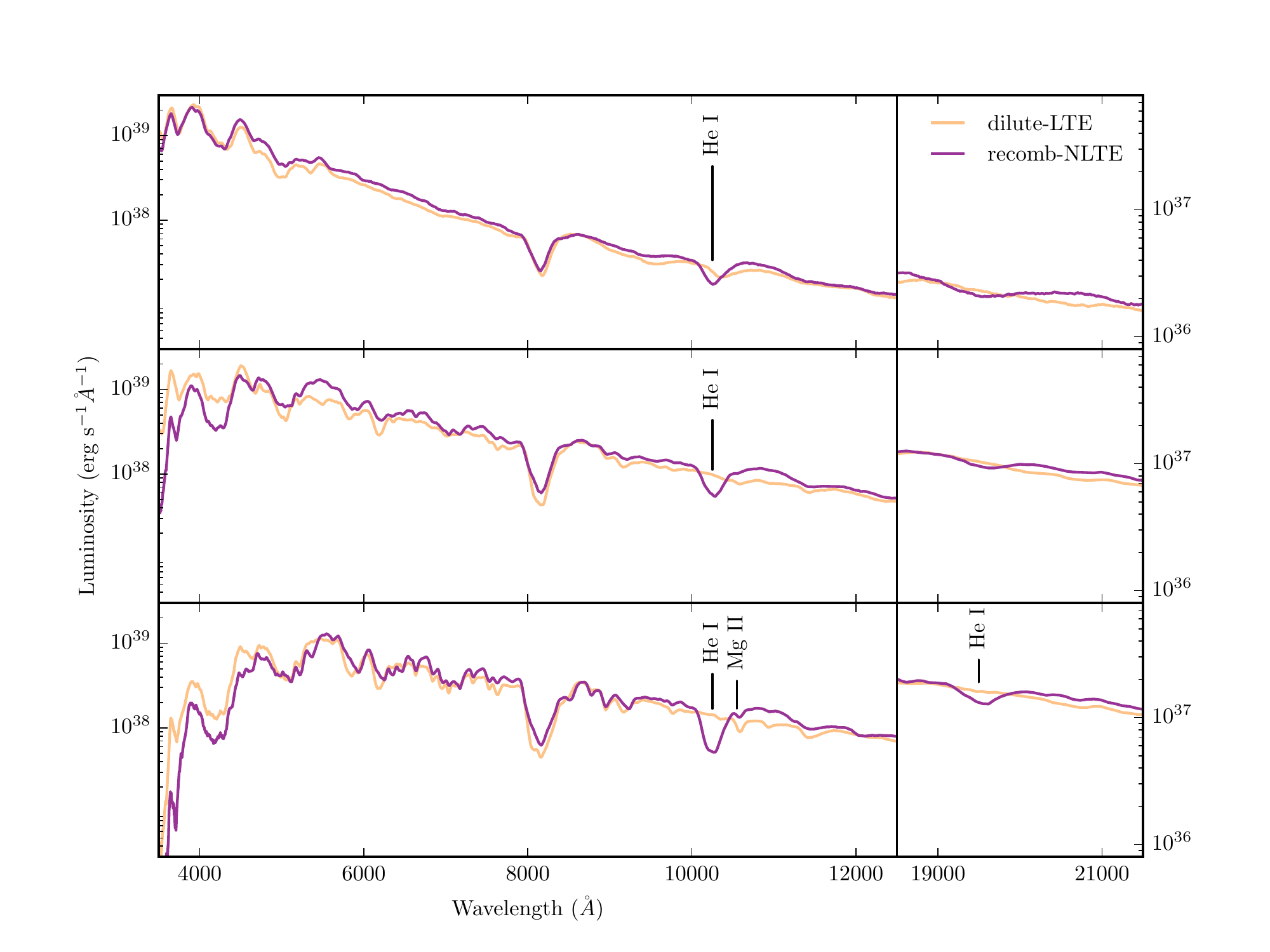}
\caption{Spectra obtained for the LM model at epochs 7, 14 and 21 days after explosion. The 10830 \AA~line is observed clearly at all epochs and grows stronger with time.}\label{fig_spectra_lm}

\end{figure*}

The spectra obtained for the LM model are shown in Figure \ref{fig_spectra_lm}. In this model the He 10830 \AA~feature is prominent at all epochs when the \texttt{recomb-NLTE} treatment is used and the 2 $\mu$m line also appears, growing stronger over time. 
As for the HM model, we also note that the treatment of He excitation/ionization has a clear impact on the strengths of other features in the spectra.

\section{Discussion}
\label{Sec_Discussion}

\subsection{Prospects for observing He~{\sc i} 10830~\AA~in double-detonation models}

\begin{figure}

\includegraphics[width=0.5\textwidth]{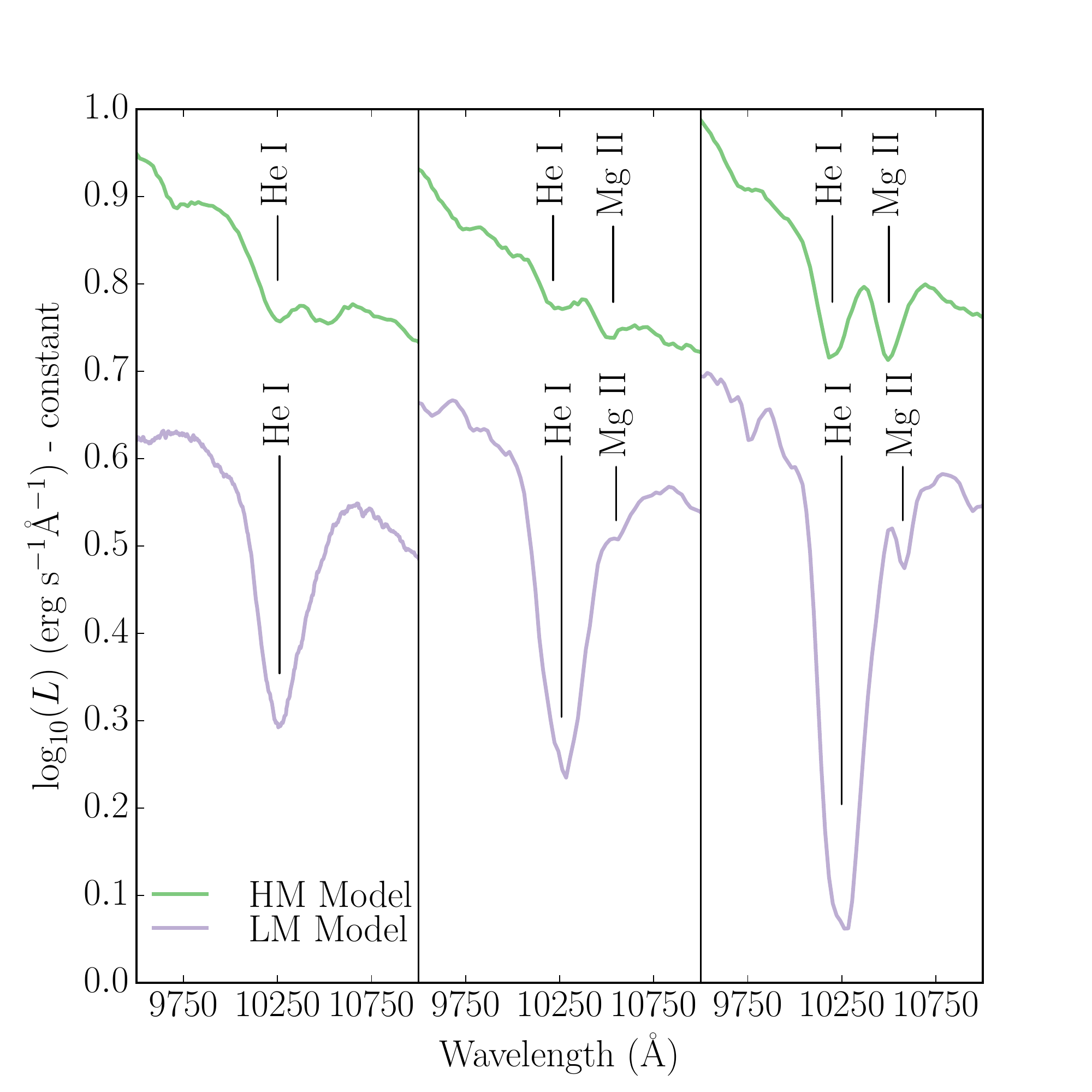}
\caption{Zoom-in of synthetic spectra around the He~{\sc i} 10830~\AA-feature for both the HM and LM models at the three epochs we consider (left panel: 7 days before maximum; middle panel: at maximum light; right panel: 7 days after maximum.
}
\label{spectra_zoom}

\end{figure}

For both the double-detonation models we have considered, our calculations suggest that the unburnt helium in the outer ejecta could lead to moderate-to-significant optical depth in the He~{\sc i} 10830~\AA\ line at epochs around maximum light. Comparing to observations around this wavelength range is therefore an effective test of the double-detonation mechanism, even for models with relatively low-mass helium outer layers (prior to detonation, our HM model had a helium layer of only $\sim 0.055~M_{\odot}$). 

In Figure~\ref{spectra_zoom} we show our detailed line profiles calculated around the He~{\sc i} 10830~\AA\ region. \cite{marion2009} have presented a substantial sample of SNe~Ia spectra covering this wavelength range across a range of epochs that includes the phases that we have considered here (see, e.g., their figure 20). They do not report any clear He~{\sc i} detections and their spectra do not show any obvious features around 10250~\AA\ (i.e. just to the blue of the Mg~{\sc ii} feature) where our models predict the He~{\sc i} 10830~\AA\ will manifest. We note, however, that the weakness of the predicted feature in the HM model would likely preclude its detection at the signal-to-noise level of some of their data. In fact, potential detections of features to the red of the Mg~{\sc ii} line have been reported based on high-quality data of a few SNe~Ia. Specifically, features were identified in the subluminous 91bg-like SN~Ia 1999by \citep{hoeflich2002} and the underluminous SNe~Ia iPTF13ebh \citep{hsiao2015} that appear clearly separated from the Mg~{\sc ii} profile while evidence of an additional feature affecting the blue wing of the Mg~{\sc ii} profile has been reported for the normal SNe~Ia 2011fe \citep{hsiao2013} and 2014J \citep{marion2015} (see e.g. figure~8 of \citealt{hsiao2015}). In all these cases, it has been proposed that the detected feature is associated with C~{\sc i} 10693~\AA~at a velocity of $\sim 12,000$ -- $15,000$~km~s$^{-1}$. However, our calculations suggest a potential alternative/additional contribution, namely that this feature may be associated with the high velocity He~{\sc i} 10830~\AA~line in double detonation models: the velocities predicted for the largest opacities in He~{\sc i} 10830~\AA~by the HM model ($\sim 19,000$ km~s$^{-1}$, see Figure~\ref{fig_tau_sobolevs_hm}) lead to a wavelength coincidence with the C~{\sc i} 10693~\AA~identifications for lower velocity. 

In the LM model, the He~{\sc i} 10830~\AA~feature is predicted to be much more prominent and is clearly too strong to be consistent with the spectra of the normal SNe~Ia 2011fe or 2014J (see e.g. figure~8 of \citealt{hsiao2015}). However, owing to its low $^{56}$Ni mass ($\sim 0.2~M_{\odot}$ in total, combining core and shell detonations), the LM model spectra are more relevant for guiding the interpretation of significantly sub-luminous events. It is therefore interesting to note that the spectra of 1999by \citep{hoeflich2002} do show a relatively prominent feature around 10250~\AA~(hitherto identified as C~{\sc i}). Although this is clearly not as strong as the He~{\sc i} 10830~\AA-line predicted by our LM model, the wavelength/velocity correspondence is good and the qualitative trend towards a stronger feature in a less luminous explosion is consistent with our modelling.

Further refinement of the models is needed to aid quantitative comparison with observations, but overall our results clearly suggest that there are good prospects for constraining double-detonation models via the high-velocity He~{\sc i} 10830~\AA~line and that consideration should be given to the potential for this line to affect interpretation of the spectral region around Mg~{\sc ii} 10927~\AA~ and/or C~{\sc i} 10693~\AA. Particular emphasis may be placed on constraining the applicability of double-detonation models to faint thermonuclear transients, where our calculations clearly suggest that the  He~{\sc i} 10830~\AA~line can be strong.

Our calculations suggest that the He~{\sc i} 2 $\mu$m line may also be an effective diagnostic at similar epochs for some models: this line does become strong in our LM model, but it does not appear clearly in our HM model, where the He density is higher.

\subsection{Effects of Helium NLTE Treatment on Other Elements}
\label{Sec_Discuss_Other}

As noted in Section~\ref{Sec_Results}, the 
use of alternative methods for treating helium has indirect influences on the spectral features of other elements. This can be understood as a consequence of the difference in the free electron density (and therefore ionization balance) in the outer ejecta that arises from the treatment of helium: under the \texttt{dilute-LTE}/\texttt{nebular} approximations, helium is mostly neutral (leading to a low electron density) while the \texttt{recomb-NLTE} posits that helium is ionized by non-thermal processes, which results in a substantial increase in the number density of free electrons. 
A prominent example is the change in the high-velocity Ca line around 8000 \AA~in Figure \ref{fig_spectra_hm}. This is a consequence of an increased Ca {\sc ii} population (at the expense of Ca {\sc iii}), driven by the higher free electron density in the \texttt{recomb-NLTE} calculation. It is particularly noteworthy that this indirect effect of the treatment of He on the blue wing of the Ca~{\sc ii} profile is dramatic even at epochs when the He~{\sc i} 10830~\AA~ line is extremely weak (see middle panel of Figure~\ref{fig_spectra_hm}). 
Although the effect of the NLTE treatment of helium on the states of other elements is obviously complex, it does provide an interesting prospect that the presence of helium may affect high-velocity components in double-detonation models, even when no helium lines are directly identifiable. We caution, however, that this study has focused only on helium and that further work that considers more complete treatment of the ionization (including non-thermal ionization) and recombination of heavier elements would be needed to draw firm conclusions.

\subsection{Comparison with previous spectral modelling}\label{Sec_Comparison}

A number of previous studies have made predictions for synthetic observables of double-detonation models \citep[e.g.][]{hoeflich1996, nugent1997, kromer2010, woosleykasen2011}. In particular, 
\cite{nugent1997} discuss the formation of He~{\sc i} lines in the double detonation model of \cite{woosleyweaver1994}, which is based on a progenitor CO core of 0.7~$M_{\odot}$ surrounded by a thick He shell of 
0.2~$M_{\odot}$ at ignition, and has luminosity comparable to our HM model.
Although \cite{nugent1997} report that no clear signatures of He~{\sc i} (or He~{\sc ii}) manifest in their synthetic optical spectra, they do show substantial optical depths for the near-IR He~{\sc i} 10830~\AA~line at some points in the model (up to $\tau \sim 10^4$, see their figure 7). Our computed 10830~\AA~optical depths never reach such high values (see Figures~\ref{fig_tau_sobolevs_hm} and \ref{fig_tau_sobolevs_lm}), but our synthetic spectra still suggest that the 10830~\AA~ line could be detectable, even in our HM model (despite its considerably lower He shell mass).

Following observations of SN1994D reported by \cite{meikle1996}, \cite{mazzali1998} studied the formation of He~{\sc i} features (including the 10830~\AA~line) in models for SNe~Ia.  By introducing helium into the W7 model of \cite{nomoto1984}, they showed it was possible to make the 10830~\AA~line appear in their synthetic spectrum: depending on how the helium was introduced, they estimated that a helium mass of 0.014 - 0.03~M$_{\odot}$ was needed to produce a feature of comparable strength to that for which \cite{meikle1996} has suggested the He~{\sc i} identification in SN1994D. These He masses are similar to the shell He masses in the models considered here and thus our results are in line with the findings of \cite{mazzali1998}: such small masses can make clear He~{\sc i} lines. However, our models do not support a plausible identification of the feature reported by \cite{meikle1996} with He~{\sc i} in double detonation models: our models predict He~{\sc i} 10830~\AA\ velocities that are considerably too high ($\sim 16,000 - 18,000$km~s$^{-1}$ compared to $\simlt 13,000$km~s$^{-1}$) to match the features in SN1994D \citep{meikle1996,mazzali1998}.

\cite{dessart2014} performed NLTE radiative transfer simulations for a model of a He shell detonation on the surface of a CO WD. Such a model is closely related to the double-detonation models we have considered, the key difference being that the secondary detonation of the CO core is not invoked. 
The structure of the model ejecta in our double-detonation model is, of course, quite different to the case of pure He shell detonation (compare fig. 1 of \citealt{dessart2014} to Figures~\ref{fig_abundances_hm} and \ref{fig_abundances_lm}; see also \citealt{sim2012}). Nevertheless, our findings on He~{\sc i} in double-detonation models are broadly in line with those of \cite{dessart2014}:
they found that the 10830~\AA~He~{\sc i} was prominent for a range of epochs around maximum light (in their case, $\sim 4 - 20$~days post-explosion; note that the light curve evolution is relatively fast for their shell-only detonation) and that the 2 $\mu$m He~{\sc i} feature became increasingly strong with time. 
This is quite consistent with our findings for the LM model, which has a helium mass ($0.077~M_{\odot}$) similar to, but a little smaller than, their model ($0.112~M_{\odot}$). There is, however, one interesting difference, which in principle allows the models to be distinguished: in the double-detonation models, the ash from the He shell is entirely located at high velocities (the low velocity ejecta being dominated by CO detonation ash: see Figures~\ref{fig_abundances_hm} and \ref{fig_abundances_lm}); in contrast, the shell-only detonation leads to ejecta with helium present at all velocities (see fig 1 of \citealt{dessart2014}). Consequently, the He~{\sc i} 10830~\AA~line profile obtained in our double-detonation model is clearly detatched: the absorption part of the profile is all strongly blueshifted, and the emission is correspondingly broad (and shallow). In contrast, the shell-only detonation model leads to a 10830~\AA~profile that extends down to low velocities and is more reminiscent of the standard P~Cygni shape.

We note that, at the latest epochs they model, \cite{dessart2014} find that the He~{\sc i} 2 $\mu$m line is strong and in emission. In future work it will be interesting to investigate how important this feature becomes in the late phases of the double-detonation models but, owing to the physical simplifications currently made in \textsc{tardis}, we cannot extend this study to such late epochs.

\subsection{Validity of the New Approximation}

By comparison to the results obtained using the full statistical equilibrium code of Hachinger et al. (2012), our newly-developed approximation, \texttt{recomb-NLTE}, has been shown to be successful in describing the excitation state of key He~{\sc i} levels in the regime of interest to this study. 
The original justification for treating the He~{\sc i} excited states as in dilute LTE with the He~{\sc ii} ground state was based on the reasoning and results of \cite{hachinger2012}, as outlined in Sections \ref{Sec_Intro} and \ref{Sec_New_Approx}, and this assumption has proved to be successful in our own models. The unique treatment of the metastable states was justified theoretically by the fact that these states are populated more strongly by de-excitation from higher levels than direct recombination, specifically by their respective 2P states due to the high oscillator strengths of the associated transitions. This reasoning is supported empirically by the extraction and analysis of the recombination and de-excitation rates from the rate-equation solver of \cite{hachinger2012} for test cases from Figure \ref{fig_level_populations_hm}.

Since the two models investigated have quite different parameters, it seems that this approximation could be applicable (with caution) to a wide range of double-detonation models, partly alleviating the need for full numerical NLTE treatment of helium during explorations of parameter space.
One important limiting factor for the use of the \texttt{recomb-NLTE} approximation is whether the non-thermal ionization rate is sufficiently high that helium remains ionized. The appearance of the key helium features emphasised in this paper is primarily dependent on the populations of the two lowest excited states. The fractional accuracy of these populations is effectively equivalent to the level of deviation of the helium population from full ionisation, provided that these excited state populations remain most strongly coupled to the He {\sc ii} ground state.
Thus in cases where a significant degree of ionisation is expected but the true ionisation state cannot be estimated, it is reasonable to use the approximation to determine an approximate upper limit on the potential strength of helium features for a particular model. 

\section{Summary and Conclusion}
\label{Sec_Conclusions}

The primary aim of this work was to determine whether the current generation of double detonation models that involve relatively low-mass He shells could conceal the presence of helium in their photospheric-phase spectra, even when non-thermal effects on the state of helium are approximately taken into account. 
We focused on two potential models: first, a high-mass model with luminosity appropriate for normal SNe Ia, and second a lower mass model that has been presented as a candidate for less luminous thermonuclear transients.

In the high-mass model, we found that the He~{\sc i} 2 $\mu$m line optical depth is likely to remain too low for the feature to be detected in absorption at the phases we consider (within a week of maximum light). However, we found that the He~{\sc i} 10830~\AA-line may have a moderate optical depth in the high-velocity helium-rich outer ejecta and may therefore be observable at photospheric epochs around (and after) maximum light. Since the helium is located in a high-velocity shell, the predicted 10830~\AA~profile is detached and substantially blueshifted ($\sim 19,000$~km~s$^{-1}$): specifically, our calculations for the high-mass predict that the deepest absorption will be at around $\sim 10250$~\AA, a region where the feature may be blended with the blue wing of Mg~{\sc ii} 10927~\AA~ or confused/blended with C~{\sc i} 10693~\AA. Given the recent identifications of features with C~{\sc i} 10693~\AA~in high-quality observations of normal (and subluminous) SNe~Ia \citep{hsiao2013,marion2015,hsiao2015}, the possible confusion of this feature with He~{\sc i} and the potential ramifications, both for the understanding of signatures of double-detonation and of unburned carbon, warrant further detailed investigation.

The mass of the helium shell adopted in our high-mass model is close to the minimum shell mass that was estimated for helium detonation around a $\sim 1~M_{\odot}$ CO WD by \cite{bildsten2007}. Thus our calculations suggests that, even for minimum shell masses, the 10830~\AA-feature may be potentially observable in double-detonations scenarios for normal SNe~Ia that invoke spontaneous detonation of a stably-accreted helium shell. Consequently, it may be challenging to ``conceal'' the presence of helium in the near-IR spectra of such models (other than by line blending).
However, further investigation of other scenarios is needed. In particular, \cite{shenmoore2014} show that helium-rich outer shells with masses below 0.01~$M_{\odot}$ may still support steady detonations for 1~$M_{\odot}$ CO cores and it has been argued that ignition of very low mass surface helium shells may be triggered during dynamical mergers (e.g. \citealt{guillochon2010}, \citealt{raskin2012}, \citealt{pakmor2013}) and potentially lead to CO core detonation \citep{pakmor2013}.
In such models the high-velocity helium mass could be further reduced by a factor of a few compared to the model we consider here, making detection of the He~{\sc i} even more difficult.
 
In the lower mass model we consider, the helium features at 10830~\AA~and 2 $\mu$m are both predicted to be strong and could provide a clear means to identify the presence of helium in the explosion. This is qualitatively comparable to the results of the shell-only helium detonation model studied by \cite{dessart2014} but there are important differences in the helium features predicted at photospheric epochs: specifically, the secondary CO detonation included here has a clear impact on the He~{\sc i} line profile, potentially allowing the two scenarios to be distinguished spectroscopically. Thus attempts to observe the He~{\sc i} near-IR features in sub-luminous transients for which helium detonation models are considered \citep[e.g.][]{inserra2014} could help not only to confirm the presence of helium but to determine explosion mechanisms.

In the process of this work, we have developed a new analytic approximation that can be used to estimate the excited He~{\sc i} level populations in double-detonation supernova models. This captures key atomic physics related to helium without incurring significant computational cost, and it should allow for simulations of this type to be carried out relatively simply in the future.

\section*{Acknowledgments}
We are grateful to K. Maguire, M. Kromer and R. Kotak for useful discussions related to this work. AB acknowledges support from a summer studentship with the Astrophysics Research Centre at Queen's University Belfast during which part of this work was carried out, and for support via the 2015 Google Summer of Code program. SAS acknowledges support from the STFC grant ST/L000709/1.WEK acknowledges the support from the ESO Fellowship at the European Southern Observatory. We thank the anonymous referee for comments that helped to clarify the paper.

\bibliographystyle{mn2e}

\begin{thebibliography}{99}
\bibitem[Bildsten et al.(2007)]{bildsten2007} Bildsten, L., Shen, 
K.~J., Weinberg, N.~N., \& Nelemans, G.\ 2007, ApJ, 662, L95
\bibitem[Cann 
\& Thakkar(2002)]{cannthakkar2002} Cann, N.~M., \& Thakkar, A.~J.\ 2002, Journal of Physics B Atomic Molecular Physics, 35, 421 
\bibitem[Chugai(1987)]{chugai1987b} Chugai, N.~N.\ 1987, Soviet Astronomy Letters, 13, 282 
\bibitem[Dere et 
al.(1997)]{dere1997} Dere, K.~P., Landi, E., Mason, H.~E., Monsignori Fossi, B.~C., \& Young, P.~R.\ 1997, A\&AS, 125, 149 
\bibitem[Dere et 
al.(2009)]{dere2009} Dere, K.~P., Landi, E., Young, P.~R., et al.\ 2009, A\&A, 498, 915 
\bibitem[Dessart 
\& Hillier(2014)]{dessart2014} Dessart, L., \& Hillier, D.~J.\ 2014, arXiv:1411.7397 
\bibitem[Drake 
\& Morton(2007)]{drakemorton2007} Drake, G.~W.~F., \& Morton, D.~C.\ 2007, ApJS, 170, 251
\bibitem[Dunkley et al.(2013)]{dunkley2013} {Dunkley}, S.~D., {Sharpe}, G.~J. \& {Falle}, S.~A.~E.~G.\ 2013, MNRAS, 431, 3429
\bibitem[Fernley et al.(1987)]{fernley1987} Fernley, J.~A., Seaton, 
M.~J., 
\& Taylor, K.~T.\ 1987, Journal of Physics B Atomic Molecular Physics, 20, 6457 
\bibitem[Fink et 
al.(2007)]{fink2007} Fink, M., Hillebrandt, W., R{\"o}pke, F.~K.\ 2007, A\&A, 476, 1133
\bibitem[Fink et 
al.(2010)]{fink2010} Fink, M., R{\"o}pke, F.~K., Hillebrandt, W., et al.\ 2010, A\&A, 514, A53
\bibitem[Fink et al.(2014)]{fink2014} Fink, M., Kromer, M., 
Seitenzahl, I.~R., et al.\ 2014, MNRAS, 438, 1762 
\bibitem[Foley et al.(2009)]{foley2009} Foley, R.~J., Chornock, 
R., Filippenko, A.~V., et al.\ 2009, AJ, 138, 376 
\bibitem[Foley et al.(2013)]{foley2013} Foley, R.~J., Challis, 
P.~J., Chornock, R., et al.\ 2013, ApJ, 767, 57 
\bibitem[Graham(1988)]{graham1988} Graham, J.~R.\ 1988, ApJ, 
335, L53 
\bibitem[Forcada et al.(2006)]{forcada2006} Forcada, R., 
Garcia-Senz, D., 
\& Jos{\'e}, J.\ 2006, International Symposium on Nuclear Astrophysics - Nuclei in the Cosmos, 96 
\bibitem[Forcada(2007)]{forcada2007} Forcada, R.\ 2007, Supernovae: 
lights in the darkness, 12 
\bibitem[Guillochon et al.(2010)]{guillochon2010}
{Guillochon}, J., {Dan}, M., {Ramirez-Ruiz}, E. \& {Rosswog}, S.\ 2010, ApJ, 708, L64
\bibitem[Hachinger et al.(2012)]{hachinger2012} Hachinger, S., 
Mazzali, P.~A., Taubenberger, S., et al.\ 2012, MNRAS, 422, 70 
\bibitem[Hillebrandt 
\& Niemeyer(2000)]{hillnie2000} 
Hillebrandt, W., \& Niemeyer, J.~C.\ 2000, ARA\&A, 38, 191
\bibitem[H\"oflich et al.(2002)]{hoeflich2002} 
{H{\"o}flich}, P., {Gerardy}, C.~L., {Fesen}, R.~A. \& 
	{Sakai}, S.\ 2002, ApJ, 568, 791
\bibitem[H\"oflich \& Khokhlov(1996)]{hoeflich1996} H\"oflich, P., \& Khokhlov, A.\ 1996, ApJ, 457, 500 
\bibitem[Hsiao et al.(2013)]{hsiao2013}
{Hsiao}, E.~Y., et al.\ 2013, ApJ, 766, 72
\bibitem[Hsiao et al.(2015)]{hsiao2015}
{Hsiao}, E.~Y., et al.\ 2015, A\&A, 578, 9
\bibitem[Hummer 
\& Storey(1998)]{hummerstorey1998} Hummer, D.~G., \& Storey, P.~J.\ 1998, MNRAS, 297, 1073
\bibitem[Inserra et al.(2014)]{inserra2014} Inserra, C., Sim, 
S.~A., Wyrzykowski, L., et al.\ 2014, arXiv:1410.6473 
\bibitem[Jordan et al.(2012)]{jordan2012} Jordan, G.~C., IV, 
Perets, H.~B., Fisher, R.~T., \& van Rossum, D.~R.\ 2012, ApJ, 761, LL23 
\bibitem[Kasliwal et al.(2010)]{kasliwal2010} Kasliwal, M.~M., 
Kulkarni, S.~R., Gal-Yam, A., et al.\ 2010, ApJ, 723, L98 
\bibitem[Kerzendorf 
\& Sim(2014)]{kerzendorf2014} Kerzendorf, W.~E., \& Sim, S.~A.\ 2014, MNRAS, 440, 387
\bibitem[Kilic et al.(2014)]{kilic2014} Kilic, M., Hermes, J.~J., 
Gianninas, A., et al.\ 2014, MNRAS, 438, L26 
\bibitem[Kromer 
\& Sim(2009)]{kromer2009} Kromer, M., \& Sim, S.~A.\ 2009, MNRAS, 398, 1809 
\bibitem[Kromer et al.(2010)]{kromer2010} Kromer, M., Sim, S.~A., 
Fink, M., et al.\ 2010, ApJ, 719, 1067
\bibitem[{\L}ach 
\& Pachucki(2001)]{lachpachucki2001} {\L}ach, G., \& Pachucki, K.\ 2001, Phys. Rev. A., 64, 042510 
\bibitem[Landi et al.(2013)]{landi2013} Landi, E., Young, P.~R., 
Dere, K.~P., Del Zanna, G., \& Mason, H.~E.\ 2013, ApJ, 763, 86 
\bibitem[Li et al.(2011)]{li2011} Li, W., Leaman, J., 
Chornock, R., et al.\ 2011, MNRAS, 412, 1441 
\bibitem[Livne(1990)]{livne1990} Livne, E.\ 1990, ApJ, 354, L53
\bibitem[Livne 
\& Glasner(1990)]{livneglasner1990} Livne, E., \& Glasner, A.~S.\ 1990, ApJ, 361, 244
\bibitem[Lucy(1991)]{lucy1991} Lucy, L.~B.\ 1991, ApJ, 383, 308 
\bibitem[Lucy(1999)]{lucy1999} Lucy, L.~B.\ 1999, A\&A, 345, 211 
\bibitem[Lundqvist et al.(2013)]{lundqvist2013} Lundqvist, P., 
Mattila, S., Sollerman, J., et al.\ 2013, MNRAS, 435, 329 
\bibitem[Marion et al.(2009)]{marion2009}
{Marion}, G.~H., {H{\"o}flich}, P., {Gerardy}, C.~L., 
	{Vacca}, W.~D., {Wheeler}, J.~C. \& {Robinson}, E.~L.\ 2009, AJ, 138, 727
\bibitem[Marion et al.(2015)]{marion2015}
{Marion}, G.~H.\ et al.\ 2015, ApJ, 798, 39
\bibitem[Mazzali(2000)]{mazzali2000} Mazzali, P.~A.\ 2000, A\&A, 363, 705 
\bibitem[Mazzali 
\& Lucy(1993)]{mazzalilucy1993} Mazzali, P.~A., \& Lucy, L.~B.\ 1993, A\&A, 279, 447
\bibitem[Mazzali 
\& Lucy(1998)]{mazzali1998} Mazzali, P.~A., \& Lucy, L.~B.\ 1998, MNRAS, 295, 428
\bibitem[Meikle et al.(1996)]{meikle1996} 
{Meikle}, W.~P.~S. et al., MNRAS, 1996, 281, 163
\bibitem[Nomoto(1980)]{nomoto1980} Nomoto, K.\ 1980, Space Sci. Rev., 27, 563 
\bibitem[Nomoto(1982)]{nomoto1982} Nomoto, K.\ 1982, ApJ, 257, 
780 
\bibitem[Nomoto(1984)]{nomoto1984} {Nomoto}, K., {Thielemann}, F.-K. \& {Yokoi}, K., 1984, ApJ, 286, 644
\bibitem[Nugent et al.(1997)]{nugent1997} Nugent, P., Baron, E., 
Branch, D., Fisher, A., \& Hauschildt, P.~H.\ 1997, ApJ, 485, 812
\bibitem[Pakmor et al.(2013)]{pakmor2013}
{Pakmor}, R., {Kromer}, M., {Taubenberger}, S. \& {Springel}, V.\ 2013, ApJ, 770, L8
\bibitem[Perets et al.(2010)]{perets2010} Perets, H.~B., Gal-Yam, 
A., Mazzali, P.~A., et al.\ 2010, Nature, 465, 322  
\bibitem[Perlmutter et al.(1999)]{perlmutter1999} Perlmutter, S., 
Aldering, G., Goldhaber, G., et al.\ 1999, ApJ, 517, 565 
\bibitem[Phillips(1993)]{phillips1993} Phillips, M.~M.\ 1993, ApJ, 
413, L105 
\bibitem[Poznanski et al.(2010)]{poznanski2010} Poznanski, D., 
Chornock, R., Nugent, P.~E., et al.\ 2010, Science, 327, 58 
\bibitem[Ralchenko(2005)]{ralchenko2005} Ralchenko, Y.\ 2005, Memorie della Societa Astronomica Italiana Supplementi, 8, 96 
\bibitem[Ralchenko et al.(2008)]{ralchenko2008} Ralchenko, Y., Janev, 
R.~K., Kato, T., et al.\ 2008, Atomic Data and Nuclear Data Tables, 94, 603
\bibitem[Raskin et al.(2012)]{raskin2012}
{Raskin}, C., {Scannapieco}, E., {Fryer}, C., {Rockefeller}, G. \&
	{Timmes}, F.~X.\ 2012, ApJ, 746, 62
\bibitem[Rauch 
\& Deetjen(2003)]{rauchdeetjen2003} Rauch, T., \& Deetjen, J.~L.\ 2003, Stellar Atmosphere Modeling, 288, 103 
\bibitem[Riess et al.(1998)]{riess1998} Riess, A.~G., Filippenko, 
A.~V., Challis, P., et al.\ 1998, AJ, 116, 1009 
\bibitem[Ruiter et al.(2011)]{ruiter2011} Ruiter, A.~J., 
Belczynski, K., Sim, S.~A., et al.\ 2011, MNRAS, 417, 408
\bibitem[Ruiter et al.(2013)]{ruiter2013} Ruiter, A.~J., Sim, 
S.~A., Pakmor, R., et al.\ 2013, MNRAS, 429, 1425
\bibitem[Ruiter et al.(2014)]{ruiter2014} Ruiter, A.~J., 
Belczynski, K., Sim, S.~A., Seitenzahl, I.~R., 
\& Kwiatkowski, D.\ 2014, MNRAS, 440, L101 
\bibitem[Savitzky 
\& Golay(1964)]{savitzkygolay} Savitzky, A., \& Golay, M.~J.~E.\ 1964, Analytical Chemistry, 36, 1627 
\bibitem[Shen 
\& Bildsten(2009)]{shen2009} Shen, K.~J., \& Bildsten, L.\ 2009, ApJ, 699, 1365
\bibitem[Shen et al.(2010)]{shen2010} Shen, K.~J., Kasen, D., 
Weinberg, N.~N., Bildsten, L., \& Scannapieco, E.\ 2010, ApJ, 715, 767
\bibitem[Shen 
\& Bildsten(2014)]{shen2014} Shen, K.~J., \& Bildsten, L.\ 2014, ApJ, 785, 61
\bibitem[Shen 
\& Moore(2014)]{shenmoore2014} Shen, K.~J., \& Moore, K.\ 2014, ApJ, 797, 46
\bibitem[Sim(2007)]{sim2007} Sim, S.~A.\ 2007, MNRAS, 375, 154 
\bibitem[Sim et al.(2010)]{sim2010} Sim, S.~A., R{\"o}pke, 
F.~K., Hillebrandt, W., et al.\ 2010, ApJ, 714, L52 
\bibitem[Sim et al.(2012)]{sim2012} Sim, S.~A., Fink, M., 
Kromer, M., et al.\ 2012, MNRAS, 420, 3003
\bibitem[Sullivan et al.(2011)]{sullivan2011} Sullivan, M., 
Kasliwal, M.~M., Nugent, P.~E., et al.\ 2011, ApJ, 732, 118
\bibitem[Taam(1980a)]{taam1980a} Taam, R.~E.\ 1980, ApJ, 237, 142 
\bibitem[Taam(1980b)]{taam1980b} Taam, R.~E.\ 1980, ApJ, 242, 749  
\bibitem[Townsley et al.(2012)]{townsley2012} {Townsley}, D.~M. and {Moore}, K. and {Bildsten}, L.\ 2012, ApJ, 755, 4
\bibitem[Utrobin (1996)]{utrobin1996} Utrobin, V.~P. 1996, A\&A, 306, 219
\bibitem[Waldman et al.(2011)]{waldman2011} Waldman, R., Sauer, D., 
Livne, E., et al.\ 2011, ApJ, 738, 21 
\bibitem[\protect\citeauthoryear{Woosley 
\& Kasen}{2011}]{woosleykasen2011} Woosley, S.~E., \& Kasen, D.\ 2011, ApJ, 734, 38
\bibitem[Woosley 
\& Weaver(1994)]{woosleyweaver1994} Woosley, S.~E., \& Weaver, T.~A.\ 1994, ApJ, 423, 371 

\end{thebibliography}

\label{lastpage}

\end{document}